\documentclass[range]{ar2e}
\usepackage{cite}
\usepackage{amsfonts}
\usepackage{amsmath}
\usepackage{slashed}
\begin{document}
\input epsf.tex    

\input psfig.sty

\def\beq{\begin{equation}}
\def\eeq{\end{equation}}

\def\d{{\rm d}}


\renewcommand{\baselinestretch}{1.3}

\newcommand{\Red}{\color [rgb]{0.8,0,0}}
\def\SSG#1{{\Red [SSG: #1]}}
\newcommand{\Green}{\color [rgb]{0,0.7,0}}
\def\AK#1{{\Green [AK: #1]}}

\def\mop#1{\mathop{\rm #1}\nolimits}
\def\Vol{\mop{Vol}}
\def\Re{\mop{Re}}
\def\Im{\mop{Im}}
\def\overleftrightarrow#1{\vbox{\ialign{##\crcr
     $\leftrightarrow$\crcr\noalign{\kern-0pt\nointerlineskip}
     $\hfil\displaystyle{#1}\hfil$\crcr}}}
\def\lsim{\mathrel{\mathstrut\smash{\ooalign{\raise2.5pt\hbox{$<$}\cr\lower2.5pt\hbox{$\sim$}}}}}
\def\gsim{\mathrel{\mathstrut\smash{\ooalign{\raise2.5pt\hbox{$>$}\cr\lower2.5pt\hbox{$\sim$}}}}}

\newcommand{\diag}{{\rm diag\,}}

\jname{Annu. Rev. Nuc. Part. Sci.}
\jyear{2010}
\jvol{60}

\title{Particle Physics Implications of F-theory}

\markboth{Heckman}{Particle Physics Implications of F-theory}

\author{Jonathan J. Heckman
\affiliation{School of Natural Sciences, Institute for Advanced Study,
Princeton, NJ 08540}}

\begin{abstract}
We review recent progress in realizing Grand Unified Theories (GUTs) in a strongly
coupled formulation of type IIB string theory known as F-theory. Our main emphasis is on the
expected low-energy phenomenology of a minimal class of F-theory GUTs. We introduce the
primary ingredients in such constructions, and then present qualitative
features of GUT models in this framework such as GUT breaking,
doublet-triplet splitting, and proton decay. Next, we review proposals for realizing
flavor hierarchies in the quark and lepton sectors. We discuss possible supersymmetry breaking
scenarios, and their consequences for experiment, as well as geometrically minimal
realizations of F-theory GUTs which incorporate most of these features.
\vspace{0.5cm}
\end{abstract}

\maketitle

\setlength{\baselineskip}{18pt}

\section{Introduction}

Of the many vacua of string theory, presumably some are compatible with experiment. Even so,
not a single string based model has yet been found which satisfies all known constraints.
Extracting predictions from this vast list of possibilities requires narrowing
the search to phenomenologically promising vacua. Once suitable assumptions are imposed,
we can then ask whether the remaining vacua are compatible with observation,
and moreover, what new phenomena to expect. Correlations between seemingly
unrelated physical features then translate into predictions of the model.

For the purposes of particle physics, it is expected that gravity only
plays a significant role at energies near the Planck scale $M_{pl} \sim 10^{19}$ GeV.
As a first step in identifying promising vacua, it is thus natural to focus on a limit where the
effects of gravity decouple from the rest of particle physics. More formally, this can be phrased as the
existence of a \textit{local model} where the particle physics degrees of freedom localize
on a spacetime filling brane which can be separated from the rest of the string
compactification. This by itself is a rather mild condition which can be met in many string constructions \cite{LocalModels},
again providing a landscape of potentially interesting vacua.

As an additional criterion for selecting promising vacua, we shall also assume that
low energy supersymmetry will soon be found, and that moreover, this should be viewed
as evidence for Grand Unified Theories (GUTs). Though circumstantial, the evidence for GUT-like structures based on
$SU(4) \times SO(4)$, $SU(5)$, or $SO(10)$ gauge groups \cite{PatiSalam, GeorgiGlashow, spinorGUT}
includes the unification of the gauge coupling constants in the Minimal
Supersymmetric Standard Model (MSSM), and the intriguing fact
that the chiral matter unifies into complete GUT multiplets charged under $SU(5)$. See \cite{RabyReview} for a recent
review of some aspects of GUTs.

Though the requirements of a local model and GUT-like structures can individually be met, combining
them turns out to significantly narrow the available options in string based models. A
strongly coupled formulation of IIB string theory known as F-theory \cite{VAFA, Morrison:1996na, Morrison:1996pp}
provides a potentially promising starting point for realizing GUTs in local models. For recent work on F-theory GUTs,
see for example \cite{DW} -- \cite{Grimm:2009yu}. Here we investigate what types of F-theory GUT geometries realize elements of
the Standard Model, and conversely, how the geometry constrains the available model building options.

\subsection{GUT-Like Structures}

We now briefly review the aspects of GUTs we shall aim to realize in string theory.
Throughout, we work in conventions compatible with $\mathcal{N} = 1$ supersymmetry, so that
all fermions are taken to be left-handed Weyl spinors.

In the MSSM, there are three generations of chiral superfields given by Standard Model analogues of the quark doublets $Q$, anti-ups $U$
and anti-downs $D$ transforming in the respective $SU(3)_{C} \times SU(2)_{L} \times U(1)_{Y}$ representations
\begin{equation}
Q : (3,2)_{1/6}, \text{\,\,} U : (\overline{3},1)_{-2/3}, \text{\,\,} D : (\overline{3},1)_{1/3}
\end{equation}
and three generations of lepton doublets $L$, and anti-leptons $E$ in the representations
\begin{equation}
L : (1,2)_{-1/2}, \text{\,\,} E : (1,1)_{1}.
\end{equation}
Compared with the Standard Model, the Higgs sector is extended to include
Higgs up and Higgs down chiral superfields in the representations
\begin{equation}
H_u : (1,2)_{1/2}, \text{\,\,} H_d : (1,2)_{-1/2}.
\end{equation}
The Higgs fields couple to the chiral matter through the superpotential terms
\begin{equation}\label{MSSMspot}
W_{MSSM} \supset \lambda^{u}_{ij} H_u Q^i U^j + \lambda^{d}_{ij} H_d Q^i D^j + \lambda^{l}_{ij} H_d L^i E^j
\end{equation}
which respectively induce masses for the up-type, down-type, and charged leptons of the Standard Model.

A remarkable feature of the MSSM is that when extrapolated up from the weak scale,
to one loop order the gauge coupling constants unify at a common value of
$\alpha_{GUT} = g^{2}/4 \pi \sim 1/24$ at an energy scale of $2 \times 10^{16}$ GeV. This
suggests unifying all interactions into a gauge group such as $SU(5)$. Indeed, the
chiral matter of the MSSM also unifies into the
$SU(5) \supset SU(3)_C \times SU(2)_L \times U(1)_Y$ representations
\begin{equation}
10_M \rightarrow Q \oplus U \oplus E \text{,}\,\,\, \overline{5}_{M} \rightarrow D \oplus L.
\end{equation}
The Higgs up and Higgs down fields of the MSSM fit in the $5_H$ and $\overline{5}_H$, and the interaction terms for the up and
down type quarks are respectively controlled by the interaction terms
\begin{equation}
5_{H} \times 10_M \times 10_M \,\,\,\,\text{and}\,\,\,\, \overline{5}_H \times \overline{5}_M \times 10_M.
\end{equation}
This is only the first step in realizing a GUT model. For example, a realistic GUT
must also contain mechanisms for breaking the GUT group;
removing the Higgs triplets from the $5_H$ and $\overline{5}_H$; generating more realistic
flavor textures; maintaining a sufficiently stable proton; and coupling to a supersymmetry breaking sector.

\subsection{Local Models}

Our main focus in this article will be on a class of string theory based
constructions known as \textit{local models} \cite{LocalModels}. Since gravity is quite weak
even at the GUT scale, the aim of a local model is to first reproduce details of
the Standard Model, and worry about gravity later. Geometrically, a local model
corresponds to a system where gauge theory localizes
on a $(4 + k)$-dimensional subspace of the
ten-dimensional string theory. Starting from the ten-dimensional Einstein-Hilbert
action, and the $(4+k)$-dimensional Yang-Mills action, the resulting gauge coupling constant and
four-dimensional Newton's constant respectively depend on the internal volumes as
\begin{equation}
(g^{4d}_{YM})^{2} \propto \text{Vol}(\mathcal{M}_{k})^{-1}, \,\,\,
G^{4d}_{N} \propto \text{Vol}(\mathcal{M}_{6})^{-1}.
\end{equation}
Decoupling gravity corresponds to a limit where the ratio of the characteristic
radii $(\text{Vol}(\mathcal{M}_{k}))^{1/k}/(\text{Vol}(\mathcal{M}_{6}))^{1/6}$
becomes parametrically small.

In perturbative type II superstring theory,
this suggests associating the closed string sector with gravity, and the open string sector
with gauge theory. In this case, the Standard Model degrees of freedom localize on Dirichlet- or D-branes. As a point of terminology,
we shall refer to a Dp-brane as one which fills the temporal direction and p spatial directions. In F-theory
we shall encounter non-perturbative generalizations of D-branes which allow us extra model building options. For reviews of
model building efforts with open strings, see \cite{INTDBRANES}.

Perturbative open string degrees of freedom provide nearly all of the
qualitative ingredients necessary to realize the Standard Model. The
transfer matrix of open strings between such D-branes defines the color space
degrees of freedom of the gauge theory. For example, $n$ spacetime
filling D3-branes at the same point in $\mathbb{R}^{6}$ realizes
four-dimensional $\mathcal{N} = 4$ $U(n)$ gauge theory. It is also possible to
engineer $SO$ and $USp$ gauge theories using unoriented open strings.
Open strings stretched between different stacks of D-branes correspond to
matter fields charged under the fundamental of one gauge group,
and the anti-fundamental of the other. With unoriented strings
it is also possible to generate matter charged in the two index
symmetric and anti-symmetric representations of a gauge group. Note
that all of the matter content of the Standard Model transforms in such representations.
In string perturbation theory, the gauge invariant interaction terms always
contract fundamental indices with anti-fundamental indices.

There are, however, drawbacks to local models based on purely perturbative degrees of freedom. Much of this tension stems from
the absence of certain GUT-like structures. While it is possible to engineer $SU(5)$, or more precisely $U(5)$ gauge theories with the matter content of a GUT, terms such as the $5_H \times 10_M \times 10_M$ interaction cannot be generated in string perturbation theory because such interaction terms do not match up an equal number of fundamental and anti-fundamental indices. Other GUT structures based on $SO(10)$ or exceptional groups fare even less well. The spinor $16$ of $SO(10)$ cannot be realized in open string perturbation theory, and E-type gauge group structures are also unavailable. All of this points to the need for extra non-perturbative ingredients.

\subsection{Organization of this Review}

The rest of this review is organized as follows. In section \ref{BuildingBlocks} we introduce the primary
building blocks of F-theory GUTs. Next, in section \ref{4dpuzz} we review how this class of models addresses some of the qualitative
issues present in four-dimensional GUT models. In section \ref{SUSYbreak} we
discuss supersymmetry breaking scenarios. Section \ref{FlavorFth} discusses
more detailed aspects of flavor physics for quarks and leptons, and in section \ref{eunif}
we review some of the characteristics of minimal models based on a single point of $E_8$ unification.
Section \ref{Conclude} presents our conclusions and possible directions for future investigation.

The intent of this article is to provide a point of entry to the literature. For earlier reviews with
slightly different emphasis, see \cite{HVLHC, WijnholtReview, BourjailyReview, Vafa:2009se}. In addition, some important topics will
only be treated briefly, if not at all. Most noticeably absent is a discussion of
recent efforts to include the effects of gravity \cite{Blumenhagen:2008zz, Andreas:2009uf, DWIII,
Blumenhagen:2009up, Marsano:2009ym, Marsano:2009gv, Blumenhagen:2009yv, Marsano:2009wr, Grimm:2009yu}.

\section{Building Blocks of F-theory GUTs} \label{BuildingBlocks}

\subsection{Brief Review of F-theory}

A remarkable feature of type IIB string theory is that it is invariant under the duality group $SL(2,\mathbb{Z})$, corresponding to all $2 \times 2$ matrices with integer entries with unit determinant. Under the action of the $SL(2,\mathbb{Z})$ duality group, the Neveu-Schwarz (NS) and Ramond Ramond (RR) two-forms transform as a two-component doublet. Moreover, complexifying the string coupling $g_s$ by combining it with the RR zero-form $C_0$, the axio-dilaton
\begin{equation}
\tau = C_0 + \frac{i}{g_s}
\end{equation}
transforms as $\tau \rightarrow (a \tau + b)/(c \tau + d)$ under integers $a,b,c,d$ such that $ad - bc = 1$.

Under the weak/strong duality map $\tau \rightarrow -1/\tau$ given by $a = d = 0$ and $b = -c = -1$, the fundamental string and D1-brane interchange roles. For more general duality group transformations the fundamental string maps to a bound state of $p$ fundamental strings and $q$ D1-branes or a $(p,q)$ string. Since fundamental strings end on D-branes, acting by the duality group provides us with new strongly coupled $(p,q)$ analogues of the more familiar D-branes on which strings of the same $(p,q)$ type can end. Branes of different $(p,q)$ type can also form non-perturbative bound states.

The profile of the axio-dilaton is closely connected with the presence of
seven-branes in the compactification. As an illustrative example, consider
a D7-brane filling $\mathbb{R}^{7,1}$ and sitting at a point of the
remaining two spatial directions. The one-form flux $F_{1}$ sourced by
the D7-brane is locally defined by the relation $F_{1} = dC_{0}$. Globally,
we cannot write $F_1$ in this form. Indeed, passing around a circle
surrounding the D7-brane, $C_0$ shifts as $C_0 \rightarrow C_0 + 1$.
Similarly, $n$ D7-branes source $n$ units of $F_1$ flux, inducing $C_0 \rightarrow C_0 + n$. This corresponds to an $SL(2,\mathbb{Z})$ transformation with $a = d= 1$, $b = n$ and $c =0$. Acting by $SL(2,\mathbb{Z})$ transformations, we see that different seven-branes will affect the axio-dilaton in different ways.

F-theory \cite{VAFA} provides a geometric formulation of strongly coupled type IIB vacua which automatically keeps track of $\tau$ in the presence of seven-branes. The main idea is to specify the profile of $\tau$ by including two additional geometric directions in addition to the usual ten spacetime
dimensions of string theory. Let us stress that these two extra dimensions are on a somewhat different footing from the other ten. In this twelve-dimensional geometry, the parameter $\tau$ is to be viewed as the shape of a two-torus, or elliptic curve \cite{VAFA}. This two-torus can be visualized as a parallelogram in the complex plane $\mathbb{C}$, with vertices at $0$, $1$, $\tau$ and $\tau + 1$, and parallel sides identified. The shape of this torus is invariant under $SL(2,\mathbb{Z})$ transformations $\tau \rightarrow (a \tau + b)/(c \tau + d)$, which makes manifest the action of the $SL(2,\mathbb{Z})$ duality group of IIB string theory. The presence of seven-branes will affect the value of $\tau$, and is indicated in the geometry by allowing $\tau$ to attain different values at different points in the spacetime.

For the purposes of model building, we will be particularly interested in
compactifications of F-theory which preserve four real supercharges in
four dimensions. Retaining this amount of supersymmetry imposes the geometric condition
that these supercharges be covariantly constant in the four complex-dimensional internal space.
Much as in other string compactifications, this condition is met for F-theory compactified on
a four complex-dimensional Calabi-Yau space. Unlike other string compactifications, the
six physical internal directions define a complex threefold $B_3$ which
need not be Calabi-Yau.

The geometry of the Calabi-Yau fourfold automatically includes the effects of seven-branes. As we have seen, encircling a seven-brane
leads to a $SL(2,\mathbb{Z})$ transformation of $\tau$. While this is an acceptable description away from the seven-brane, this means that
on top of the seven-brane, the shape of the two-torus becomes singular, and effectively pinches off. As a hypersurface
in $\mathbb{C}^2$ with coordinates $x$ and $y$, the elliptic curve can be modeled in \textit{Weierstrass form} as
\begin{equation}
y^{2} = x^{3} + f x + g
\end{equation}
where $f$ and $g$ will in general have non-trivial position dependence on $B_3$.
The location of the seven-branes are then given by roots of the discriminant of the cubic in $x$
\begin{equation}
\Delta = 4f^{3} + 27 g^2 .
\end{equation}
The equation $\Delta = 0$ corresponds to a one complex dimensional equation in $B_3$, and so
sweeps out a two complex-dimensional subspace. As a polynomial in the coordinates of
the threefold base $B_3$, $\Delta$ may factorize into irreducible polynomials so
that $\Delta = \Delta_{1} \cdots \Delta_{n}$. Each factor $\Delta_i = 0$
then defines the location of seven-branes in $B_3$.

\subsection{Geometry and Gauge Theory}\label{ssecgeo}

The locations of seven-branes are dictated by where the elliptic curve of an F-theory compactification pinches off. The precise ways that this pinching off can occur admits a classification in terms of the ADE Dynkin diagrams familiar from gauge theories, which is known as the Kodaira classification of
singular fibers. Locally, we can describe the eight real-dimensional space of the Calabi-Yau fourfold in terms of two complex directions spanned by
the seven-brane, and two additional complex directions given by an ADE type singularity (see table \ref{ADEtab}).
As their name suggests, these singularities correspond to geometries in which a number of $S^{2}$'s intersecting according to the ADE Dynkin diagram
have collapsed to zero size.

\begin{table}[h!]
   \centering
        \caption{\bf Dictionary for ADE singularities and seven-brane gauge groups.}\label{ADEtab}
        \vspace{0.3cm}
\begin{tabular}{@{}ccc@{}}%
\toprule
ADE Type & Equation in $\mathbb{C}^{3}$ & F-theory Gauge Group \\
\colrule
$A_n$               & $y^2 = x^2 + z^{n+1}$  & $SU(n + 1)$ \\
$D_n$               & $y^2 = x^2z + z^{n-1}$  & $SO(2n)$  \\
$E_6$               & $y^2 = x^3 + z^4$  & $E_6$ \\
$E_7$               & $y^2 = x^3 + xz^3$  & $E_7$ \\
$E_8$               & $y^2 = x^3 + z^5$  & $E_8$ \\
\botrule
\end{tabular}
\end{table}

In M-theory and type IIA string theory, it is known that compactifying on an ADE singularity realizes
the corresponding ADE gauge group $SU(n)$, $SO(2n)$ or $E_{6,7,8}$ in the uncompactified directions, and the
situation in F-theory is no different. For example, in M-theory, integrating the three-form potential $C_{abc}$ over
each $S^2$ of the Dynkin diagram yields a $U(1)$ gauge boson. M2-branes wrapped over
the remaining $S^{2}$'s are charged under these $U(1)$'s, and constitute the off-diagonal
gauge bosons of the ADE gauge group. The duality between circle compactifications of IIB and IIA string theory
lifts to F-theory and M-theory, allowing us to translate this derivation over to the F-theory side. Although
apparently unsuitable for GUT phenomenology, it is also possible to engineer the non-simply laced
groups $USp(2n)$, $SO(2n+1)$, $F_4$, and $G_2$.

Associated with a singular ADE fiber of an F-theory compactification is a corresponding seven-brane with that ADE gauge group.
Just as in perturbative type IIB string theory, spacetime filling seven-branes which occupy the same subspace in the internal
directions will allow us to engineer higher-dimensional gauge theories. Though our ultimate interest will be in
seven-branes which wrap four compact internal directions, let us first consider the maximally supersymmetric gauge
theory of seven-branes on $\mathbb{R}^{7,1}$ with gauge group $G_S$ engineered by an ADE singularity of type (by abuse of notation) $G_S$. To arrive at the eight-dimensional seven-brane theory in flat space, let us first consider ten-dimensional supersymmetric Yang-Mills theory on $\mathbb{R}^{9,1}$ with sixteen real supercharges and gauge group $G_S$. The field content of this ten-dimensional theory consists of an adjoint-valued gauge boson $A_{I}$, and an adjoint-valued gaugino, which transforms as a sixteen component Majorana-Weyl spinor. Dimensionally reducing this ten-dimensional theory to eight dimensions yields the maximally supersymmetric eight-dimensional gauge theory. Focussing on the bosonic field content, reduction to eight dimensions yields an eight-dimensional gauge boson transforming as a vector under $SO(7,1)$ given by those components of $A_{I}$ with directions along the eight spacetime directions. The two additional gauge bosons of $A_{I}$ then combine to form a complex scalar $\Phi$.

The eigenvalues of the complexified scalar $\Phi$ determine the positions of the seven-branes in the threefold base $B_3$. Since the order of the singularity determines the gauge group on the seven-brane, these vevs must also deform the geometry of the ADE type singularity. Consider for example a stack of $n$ seven-branes located at $z = 0$ with gauge group $SU(n)$. This corresponds to an $A_{n-1}$ singularity
\begin{equation}
y^2 = x^2 + z^n.
\end{equation}
We can break the gauge group down to $U(1)^{n-1}$ through the vev $\Phi = \text{diag}(t_1,...,t_n)$ where $t_1 + ... + t_n = 0$.
In the geometry this translates to the deformation
\begin{equation}
y^2 = x^{2} + \prod _{i = 1}^{n}(z-t_i).
\end{equation}
Note that for generic values of $z$, each factor in this product behaves roughly as a constant. However, near the locus $z = t_i$, we encounter the location of one of the seven-branes. This can be visualized as keeping the center of mass of the brane system fixed and separating the $n$ seven-branes.

More general Higgsing patterns of gauge groups also translate to complex deformations, or the \textit{unfolding} of a singularity. See \cite{KatzMorrison} for further discussion on the deformation theory of the ADE singularities. Such deformations of $G_S$ are parameterized in the gauge theory by the Casimirs of $\Phi$ which in the above example with $\Phi = \text{diag}(t_1,...,t_n)$ are given by $Tr(\Phi^{n})$, which can in turn be expressed in terms of the elementary symmetric polynomials in the $t_i$. Thus, the vevs of the Casimirs specify a breaking pattern which in turn dictates how the seven-branes are distributed in the geometry.

We can also consider more general vevs for $\Phi$ with non-trivial position dependence on the worldvolume
spanned by the original stack of seven-branes. This corresponds to rotating some of the seven-branes of the original stack so that they now occupy a different set of four internal directions of the geometry. Such seven-branes will then intersect the original stack along two real dimensions. To illustrate this, consider a $U(2)$ gauge theory described by two coincident seven-branes. When $\Phi = \text{diag}(\phi,-\phi)$, this breaks $U(2)$ down to $U(1)^{2}$. Letting $s$ denote a local holomorphic coordinate of $S$, when $\phi \sim s$, we see that for non-zero $s$, the gauge group is broken, but at $s = 0$, the $U(2)$ symmetry is restored, corresponding to the meeting of the two seven-branes. Modes $\Psi$ charged under both stacks of branes obey a wave equation obtained from the dimensional reduction of the covariant derivative of ten-dimensional Yang-Mills theory to eight dimensions. Though the full system of equations is somewhat more involved, at a schematic level, the mode $\Psi$ obeys
\begin{equation}\label{localization}
\frac{\partial \Psi}{\partial \overline{s}} - \phi \cdot \Psi = 0,
\end{equation}
so that $\Psi \sim \exp(-|s|^2)$ exhibits Gaussian falloff away from the locus of symmetry enhancement at $s = 0$. In other words, matter becomes trapped along the intersection of the seven-branes. For further discussion on matter localization in F-theory, see \cite{KatzVafa, BHVI}.

Shifting back and forth between the gauge theory description and geometric data allows us to treat some of the more subtle aspects of symmetry enhancements in an F-theory compactification in terms of eight-dimensional gauge theory. Indeed, though at generic points
the singularity type over a complex surface will be given by $G_S$, over complex one-dimensional curves, this enhancement can jump to a higher rank $G_\Sigma \supset G_S$, and over points can jump again to $G_p \supset G_\Sigma \supset G_S$. As in our discussion near equation \eqref{localization}, we can treat the system as a $G_p$ gauge theory which has been Higgsed down to the lower singularity types. The position dependence of $\Phi$ then dictates how these matter curves meet in the geometry.

\subsection{The Four-Dimensional Effective Theory}

In eight-dimensional flat space $\mathbb{R}^{7,1} = \mathbb{R}^{3,1} \times \mathbb{C}^{2}$, the components of the eight-dimensional gauge boson split up into two four-component vectors, each of which transforms as a vector under one factor of spacetime, and as a scalar of the other factor. The remaining bosonic degrees of freedom correspond to the scalar $\Phi$ which controls the position of the seven-brane. The theory of a spacetime filling seven-brane which wraps four small directions is quite similar because in a sufficiently small four-dimensional patch of the internal geometry, we can again treat the theory as an eight-dimensional gauge theory on $\mathbb{R}^{3,1} \times \mathbb{C}^{2}$. The main subtlety is how to take this local description and patch it together more globally in a way consistent with supersymmetry in $\mathbb{R}^{3,1}$. Though a full discussion would take us too far afield, the main point is that on compactifications which preserve four real supercharges, we can organize all of the field content into four-dimensional $\mathcal{N} = 1$ superfields which transform as differential forms with a given number of holomorphic and anti-holomorphic indices on the two complex-dimensional surface $S$ wrapped by the seven-brane. In terms of four-dimensional $\mathcal{N} = 1$ supermultiplets labelled by points of the internal geometry, the fields of the seven-brane theory organize into an adjoint-valued vector multiplet $W_{(0,0)}$, a chiral multiplet $\mathbb{A}_{(0,1)}$ with one anti-holomorphic index, and
a chiral multiplet $\Phi_{(2,0)}$ with two holomorphic indices \cite{DW, BHVI}.

Dimensionally reducing the eight-dimensional gauge theory to four dimensions,
the volume of the two complex-dimensional surface determines
the gauge coupling constant of the four-dimensional
gauge theory through the relation
\begin{equation}
\alpha_{GUT}^{-1} = \frac{4 \pi}{g^{2}_{YM}} \sim M_{\ast}^{4} \text{Vol}(S)
\end{equation}
where $M_{\ast}$ is a mass scale close to the string scale
$M_{\ast} \sim 10^{17}$ GeV. Roughly speaking, the radius of the volume factor
$\text{Vol}(S)^{-1/4} \sim M_{GUT}$ sets the characteristic
scale of unification.

Our discussion so far has focussed on the theory of an isolated seven-brane.
Achieving realistic phenomenology requires intersecting the GUT seven-brane
with other seven-branes of the compactification. In perturbative type IIB setups, the pairwise and triple
intersections of stacks of D7-branes respectively yield bifundamental matter and
Yukawa couplings in the four-dimensional theory. In F-theory, the analogous phenomenon
is due to other seven-branes wrapping complex surfaces $S^{\prime}$ with gauge groups
$G_{S^{\prime}}$ which intersect the GUT stack of seven-branes. In this context, the gauge theory localizes on complex two-dimensional
surfaces $S$ wrapped by seven-branes, chiral matter localizes on complex one-dimensional curves defined by the intersection of such
surfaces $\Sigma = S \cap S^{\prime}$, and cubic interaction terms such as Yukawa couplings localize at points defined by the
triple intersection of matter curves, $p = \Sigma_1 \cap \Sigma_2 \cap \Sigma_3$ (see table \ref{IngredientTab}).

\begin{table}[h!]
   \centering
        \caption{\bf Geometric Ingredients of F-theory GUTs}\label{IngredientTab}
        \vspace{0.3cm}
\begin{tabular}{@{}ccc@{}}%
\toprule
Dimension & Ingredient \\
\colrule
$10$               & Gravity \\
$8$               & Gauge Theory  \\
$6$               & Chiral Matter (+ Flux) \\
$4$               & Cubic Interaction Terms \\
\botrule
\end{tabular}
\end{table}

The intersection of seven-branes along a complex one-dimensional
curve $\Sigma$ corresponds to a further enhancement in the ADE type of the singularity, which
we denote by $G_{\Sigma} \supset G_{S} \times G_{S^{\prime}}$. As mentioned in subsection \ref{ssecgeo},
we can locally model this situation in terms of an eight-dimensional gauge theory with gauge group $G_{\Sigma}$ which is Higgsed down
to $G_{S} \times G_{S^{\prime}}$ \cite{KatzVafa, Bershadsky4d, DW, BHVI}. Some of the off-diagonal
components of the fields $\mathbb{A}_{(0,1)}$ and $\Phi_{(2,0)}$ then become trapped along the locus where the gauge group is restored
to all of $G_{\Sigma}$. The resulting theory then contains a six-dimensional hypermultiplet
charged under both $G_{S}$ and $G_{S^{\prime}}$. The adjoint representation of
$G_{\Sigma}$ decomposes into irreducible representations of $G_{S} \times G_{S^{\prime}}$, some of which
are charged under both seven-branes
\begin{equation}
ad(G_{\Sigma}) = (R,R^{\prime}) + (\overline{R},\overline{R}^{\prime}) + \cdots
\end{equation}
with $R$ and $R^{\prime}$ both non-trivial. In four-dimensional $\mathcal{N} = 1$ language,
this six-dimensional field corresponds to a collection of vector-like pairs in the
$(R,R^{\prime}) \oplus (\overline{R},\overline{R}^{\prime})$ trapped along the curve.

This recovers the usual result from perturbative type IIB string theory.
For example, the intersection of D7-branes with gauge groups $G_{S} = SU(n)$ and
$G_{S^{\prime}} = SU(m)$ contains six-dimensional matter in the $(n,\overline{m}) \oplus (\overline{n},m)$
of $SU(n) \times SU(m)$, corresponding to open strings stretched between the two stacks of D7-branes.

In an $SU(5)$ GUT, matter in the $10$ and $\overline{5}$ of $SU(5)$ are respectively
given by enhancements in the singularity type to $SO(10)$ and $SU(6)$. F-theory vacua also extend the available matter content beyond
perturbatively realized states. For example, though the spinor $16$ of $SO(10)$ cannot be
generated by open string states with two ends, it will be present in geometries where
$G_S = SO(10)$ enhances to $E_6$. This is because the decomposition of the adjoint of $E_6 \supset SO(10) \times U(1)$
contains the $16$
\begin{equation}
78 \rightarrow 45_0 \oplus 1_0 \oplus 16_{-3} \oplus \overline{16}_{+3}.
\end{equation}

The kinetic terms of the matter field wave functions are also determined by the geometry. Since these fields
localize on matter curves, in a holomorphic basis of chiral superfields, the corresponding kinetic terms
are of the form
\begin{equation}
M_{\ast}^2 \text{Vol}(\Sigma) \cdot \int d^4 \theta \Psi^\dag \Psi.
\end{equation}
In particular, matter localized on distinct matter curves will have diagonal kinetic terms. A canonical normalization of matter fields
then corresponds to performing the rescaling $\Psi \rightarrow (M_{\ast}^2 \text{Vol}(\Sigma))^{-1/2} \Psi$.

\begin{figure}[ptb]
\centerline{\psfig{figure=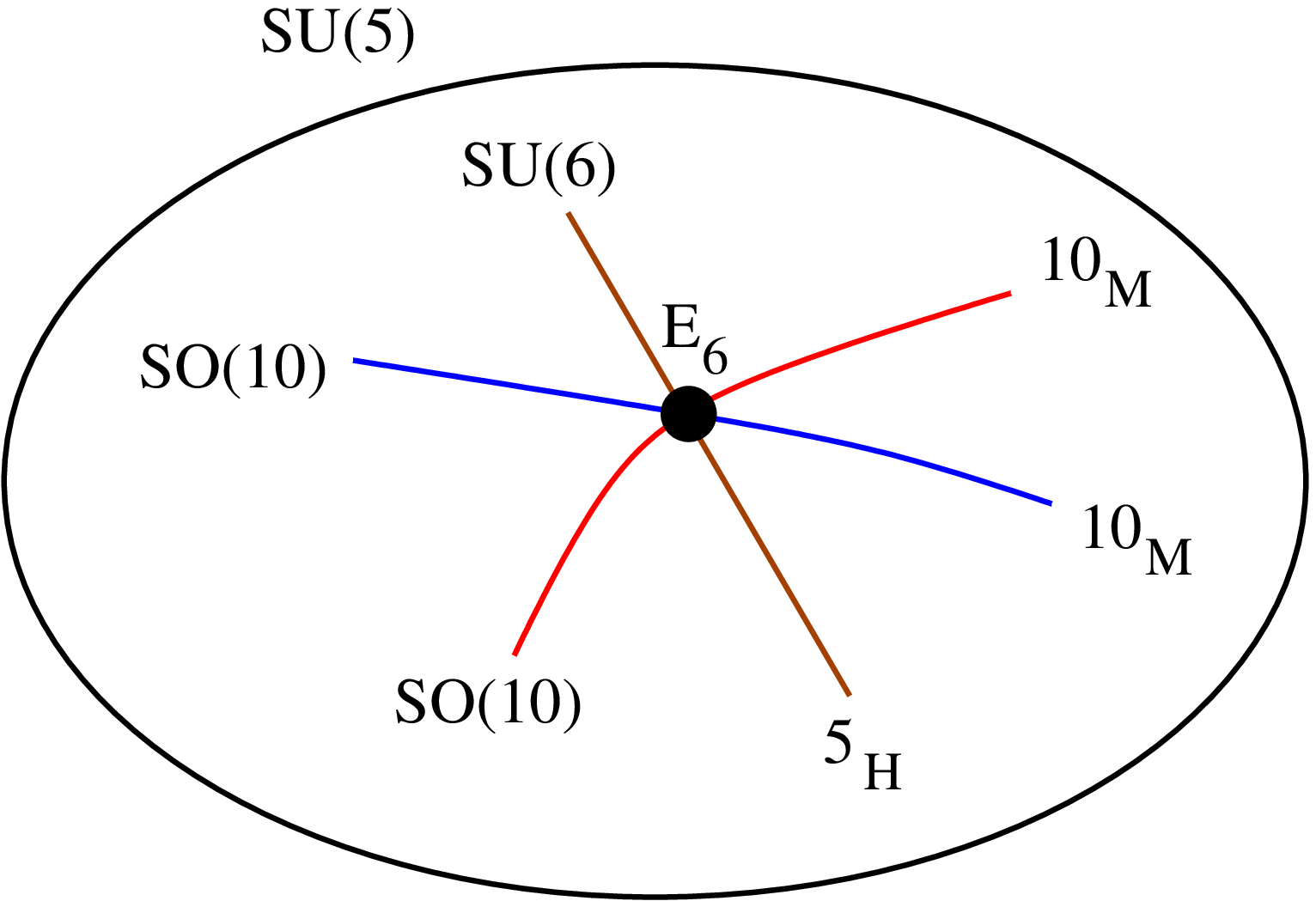,height=8cm}}
\vspace{-0.5cm}
\caption{Depiction of an $E_{6}$ singularity enhancement. The ellipsoidal
shape denotes the internal directions of the seven-brane, with gauge group
$SU(5)$. Along complex one dimensional curves, the singularity enhances to
$SO(10)$ and $SU(6)$, where matter in the $10$ and $5$ of $SU(5)$ respectively
localize. There is a further enhancement to $E_{6}$ at a point of the
geometry, where the Yukawa interaction $5_{H}\times10_{M}\times10_{M}$
localizes.}
\label{e6yuk}
\end{figure}

Further enhancements in the singularity type at points of the geometry induce interaction
terms between matter localized on curves. To illustrate this, consider a local point enhancement to
$E_6$ which is Higgsed down to $SO(10)$ and $SU(6)$ along complex curves and $SU(5)$ in the bulk of the
seven-brane. The adjoint representation of $E_6$ decomposes into irreducible representations of $SU(5) \times U(1)^2$ as:
\begin{equation}\label{tripinteraction}
78 \rightarrow 24_{0,0} \oplus 1_{0,+2} \oplus 1_{0,0} \oplus 5_{+6,0} \oplus 10_{-3,+1} \oplus 10_{-3,-1} \oplus c.c..
\end{equation}
Returning to our discussion near equation \eqref{localization}, fields with different $U(1)^2$ charges localize on different curves. An interesting feature is that in the interaction term $5 \times 10 \times 10$, each field has a distinct $U(1)$ charge so that these three matter fields automatically localize on three distinct matter curves. In other words, once two of the matter curves meet, a third comes along to form the $5 \times 10 \times 10$ interaction. See figure \ref{e6yuk} for a depiction of this geometry. Though basically correct, we will refine this statement later in section \ref{FlavorFth} when we discuss seven-brane monodromy.

More abstractly, in a neighborhood $p$ of the enhancement to a
singularity of type $G_{p}$, we can model the configuration of seven-branes in terms of an eight-dimensional theory
with gauge group $G_{p}$ Higgsed down to $G_{\Sigma_{i}}$ along various matter curves,
and $G_{S}$ elsewhere on the seven-brane. In the eight-dimensional theory, the internal gauge fields
and $\Phi_{(2,0)}$ interact through the superpotential \cite{DW, BHVI}
\begin{equation}
W_{\text{parent}} = \int Tr(\overline{\partial} \mathbb{A}_{(0,1)} + \mathbb{A}_{(0,1)} \wedge \mathbb{A}_{(0,1)}) \wedge \Phi_{(2,0)}.
\end{equation}
Expanding $\mathbb{A}_{(0,1)} = \langle \mathbb{A} \rangle + \delta \mathbb{A}$ and
$\Phi_{(2,0)} = \langle{\Phi} \rangle + \delta \Phi$ yields cubic superpotential
terms coupling the $\delta \mathbb{A}$'s and $\delta \Phi$'s of the form
\begin{equation}\label{Wpoint}
W_{p} = \lambda_{123} \Psi_{1} \Psi_{2} \Psi_{3}
\end{equation}
for chiral superfields $\Psi_{i}$ localized on curves of the internal geometry. Here, the Yukawa coupling $\lambda_{123}$ is given
by the overlap of the $\Psi_i$ wave functions in the internal geometry
\begin{equation}
\lambda_{123} = \int_{S} \psi_1 \psi_2 \psi_3.
\end{equation}
In order for the coupling of equation \eqref{Wpoint} to be generated, the product of
the $\Psi$'s must form a gauge invariant operator under the gauge groups of the seven-branes
participating in the Yukawa interaction. On the other hand, it may happen that the D-term
\begin{equation}
\int d^{4} \theta \frac{\Psi_{1}^{\dag} \Psi_2 \Psi_3}{\Lambda_{UV}}
\end{equation}
is instead gauge invariant. Integrating out Kaluza-Klein
modes of mass $\Lambda_{UV}$ localized on the $\Psi_1$ curve generate this
higher dimension operator \cite{HVGMSB}. Note that here the suppression scale $\Lambda_{UV}$ is simply the Kaluza-Klein mass scale associated with harmonics on the matter curve, which is in turn set by the characteristic radius of the seven-brane theory to be near the GUT scale. This means that if present, such terms will dominate over similar $M_{pl}$ suppressed contributions.

As an aside let us note that although these ingredients appear
to provide the most promising route for model building, it is also possible to
consider models where some of the matter content propagates in the worldvolume
of the seven-brane. In this more general context, there are
additional interactions such as those between three bulk matter
fields, and interactions between one bulk mode and two matter
fields on the same curve.

\subsection{Seven-Brane Flux}

The Standard Model has three chiral generations of matter. In an F-theory compactification,
this data is encoded in how gauge field flux in the internal directions of the seven-branes couples
to the higher-dimensional fields of the seven-brane theory.

A background flux on a seven-brane takes values in a subgroup $\Gamma \subset G_{S}$. Much
as in compactifications of the heterotic string, this breaks the gauge group in four dimensions down
to the commutator subgroup of $\Gamma$ in $G_{S}$. Note that a flux in an abelian subgroup such as
$U(1)_{Y}$ hypercharge can then break $SU(5)$ down to $SU(3)_C \times SU(2)_L \times U(1)_{Y}$.

Matter fields localized at the pairwise intersection of seven-branes with
gauge groups $G_{S}$ and $G_{S^{\prime}}$ will couple to both sets of
background fluxes. These fields obey the six-dimensional
Dirac equation
\begin{equation}
\slashed{D}_6 \Psi = (\slashed{D}_{4} + \slashed{D}_{\Sigma})\Psi = 0
\end{equation}
where $\slashed{D}_{4}$ and $\slashed{D}_{\Sigma}$ respectively denote
the Dirac operator in $\mathbb{R}^{3,1}$ and the matter curve $\Sigma$.
Since the coupling $\overline{\Psi} \slashed{D}_{\Sigma} \Psi$ of the
higher-dimensional theory descends to a mass term of the
four-dimensional theory, it follows that the massless modes correspond
to zero modes of the Dirac operator along the matter curve.

Index theory now determines the number of zero modes. For abelian field strengths
$F_{S}$ and $F_{S^{\prime}}$, the net number of chiral generations is
\begin{equation}\label{chirmodes}
\text{\# of chiral modes} = q_{S} \int_{\Sigma} \frac{F_{S}}{2 \pi} + q_{S^{\prime}}
\int_{\Sigma} \frac{F_{S^{\prime}}}{2 \pi}
\end{equation}
where the $q$'s denote the charges of the fields under the respective gauge groups. By
tuning the discrete parameters of a compactification, it is in principle
possible to realize the spectrum of the Standard Model \cite{BHVII}. Even so, as
of this writing, a fully consistent compactification realizing precisely
the matter curves and interaction terms of the Standard Model has yet to be constructed.

\section{GUT Breaking and its Consequences} \label{4dpuzz}

Having introduced the primary building blocks of an F-theory GUT,
we now study some of their consequences for low-energy physics. To frame the discussion
to follow, we shall often focus on the case of a minimal $SU(5)$ F-theory GUT with a
seven-brane wrapping a complex surface $S$. Matter in the $10_M$ and
$\overline{5}_M$ respectively localizes
on enhancements to $SO(10)$ and $SU(6)$, with Higgs fields localized
on additional $SU(6)$ enhancements. To reach the Standard Model, $SU(5)$ must be broken to
$SU(3)_C \times SU(2)_{L} \times U(1)_Y$. This has consequences for the massless spectrum,
proton decay, and gauge coupling unification.

\subsection{Local Models and GUT Breaking Fluxes}

Breaking the GUT group down to the Standard Model gauge group turns out to be surprisingly restrictive in F-theory.
In four-dimensional GUT models and higher-dimensional stringy models, the primary means to
break the GUT group are:
\begin{itemize}
\item{Vevs for adjoint-valued chiral superfields}
\item{Discrete Wilson lines.}
\end{itemize}
The local model condition obstructs both of these options in an F-theory GUT \cite{DW, BHVI, BHVII, DWII}.
In the seven-brane theory, the vevs of adjoint-valued chiral superfields determine
the location of the two complex-dimensional surface wrapped by the seven-brane $S$
inside the threefold base $B_3$. Since gravity is decoupled in a local model, this seven-brane
cannot explore $B_3$, so there are no such zero modes. As a brief aside, let us note
that in some GUT models such as flipped $SU(5)$ GUTs, GUT breaking can be realized through the vevs of fields
in representations different from the adjoint. We will briefly touch on this point later when we discuss F-theory
GUT implementations of flipped $SU(5)$ GUTs.

Wilson line breaking is also problematic because it requires a non-trivial
fundamental group $\pi_1(S) \neq 0$. Assuming there exists a limit where $B_3$
remain of finite size while $S$ contracts to a point, this requires $S$ to be
a \textit{del Pezzo} surface, namely a two complex-dimensional surface of positive curvature. For further discussion on del Pezzo surfaces geared towards physicists, see for example \cite{Iqbal:2001ye}. All such surfaces are simply connected, obstructing Wilson line breaking. Additional discussion on restrictions expected from the local model condition can be found in \cite{Andreas:2009uf, DWIII, Cordova, Grimm:2009yu}. In principle, this condition can be weakened to allow a milder version of decoupling \cite{Marsano:2009gv}.

GUT breaking by gauge field fluxes provides an alternative to these options. Recall that
non-trivial flux in a subgroup $\Gamma \subset G_{S}$ will
break the gauge group down to the commutator subgroup.
Indeed, setting $\Gamma = U(1)_{Y} \subset SU(5)$ yields
the commutator subgroup $SU(3)_{C} \times SU(2)_{L} \times U(1)_{Y}$.

It may appear surprising that this mechanism
has not been exploited in earlier work on compactifications of superstring theory. As shown in \cite{WittenNew}
this mechanism does not work in compactifications of
the perturbative heterotic string due to a generalized
Chern-Simons coupling to the Neveu-Schwarz two-form $B$ in the ten-dimensional Lagrangian density
\begin{equation}\label{HetAct}
L_{Het} \supset (dB + A \wedge \langle F \rangle)^2
\end{equation}
where $\langle F \rangle$ is the field strength in the internal directions
of the compactification. Dualizing $B$ to an axion, this generates a string
scale mass for $A$ via the St\"uckelberg mechanism. Re-mixing by other gauge bosons is possible, though this
distorts gauge coupling unification \cite{WittenNew} (see \cite{Blumenhagen:2006ux, TatarMix} for recent discussion).

The analogous coupling in the seven-brane theory is between the Ramond-Ramond
four-form potential $C_4$ and the gauge field strengths in $\mathbb{R}^{3,1}$ and the internal directions
\begin{equation}\label{genCS7bane}
\int_{\mathbb{R}^{3,1} \times S} C_{4} \wedge Tr(F_{\mathbb{R}^{3,1}} \wedge F_{S}).
\end{equation}
Assuming two of the legs of $C_4$ wedge with $F_S$, this leaves a two-form in $\mathbb{R}^{3,1}$, which is the analogue of $B$ in
equation \eqref{HetAct}.

The global topology of the complex threefold base $B_{3}$ can prevent the gauge field from developing a mass. The essential
point is that the coupling of equation \eqref{genCS7bane} involves two-forms such as $F_S$ defined purely on $S$,
and a two-form given by two legs of $C_4$ defined on the entire threefold base $B_{3}$. Depending on how $S$ embeds in
$B_3$, the integral of the wedge product of $C_{4}$ and $F_S$ can vanish. When this happens, there is no coupling
of the gauge field to an axion, and the gauge field remains massless \cite{BHVII, DWII}. This defines a mathematical condition on the relative
cohomology of $S$ and $B_3$, as well as a choice of flux $F_S$ on $S$, and it is known that it can be met in explicit examples. In practical terms, what is required is that the two-cycle Poincar\'{e} dual to $F_{S}$ in $S$ must lift to a trivial two-cycle in $B_{3}$. This is quite similar to the topological condition found in \cite{BuicanVerlinde} which prevents the $U(1)_Y$ hypercharge gauge boson from developing a mass in models based on
D3-brane probes of singularities. An interesting feature of this result is that although string dualities connect
certain heterotic and F-theoretic vacua, when this topological criterion is satisfied, no heterotic dual exists
\cite{BHVII, DWII}.

\subsubsection{Spectrum Constraints}

The presence of a background internal gauge field flux in the hypercharge direction or \textit{hyperflux} also affects the zero mode spectrum of the theory. Consider again a minimal $SU(5)$ GUT model. The internal gauge bosons of $SU(5)$ in the
$(3,2)_{-5/6} \oplus (\overline{3},2)_{5/6}$ of $SU(3)_C \times SU(2)_L \times U(1)_Y$ are in exotic representations of the Standard Model gauge group,
and can only be avoided for an essentially unique choice of cohomology class for the
hyperflux \cite{BHVII}. An important subtlety for this type of GUT breaking flux is the overall factor of
$5$ in the hypercharge of these off-diagonal elements. To actually avoid introducing exotics, it is
necessary to allow seemingly fractional hyperfluxes \cite{BHVII}. In order for all matter fields
to couple to an integral number of net flux quanta, this also requires fluxes to be activated on the other seven-branes
of the compactification \cite{BHVII, DWII}.

Increasing the rank of the GUT group only adds to the number of constraints which a background
flux must satisfy. In \cite{BHVII} it was found that breaking $SO(10)$ to $SU(3)_C \times SU(2)_L \times U(1)_Y \times U(1)$
by fluxes always generates exotics propogating in the bulk of the seven-brane. There do, however,
exist fluxes which break $SU(6)$ down to the Standard Model without introducing such bulk exotics
\cite{Chung:2009ib}.

It is also possible to construct hybrids which incorporate flux breaking to a four-dimensional model, and then
a further breaking of the GUT group down to the Standard Model. For example, in a flipped $SU(5) \times U(1)$ GUT,
a vev for matter in the $10_{-1} \oplus \overline{10}_{+1}$ leads to the Standard Model gauge group.
Partial flux breaking of $SO(10)$ down to $SU(5) \times U(1)$ has been used as a starting point for
constructing flipped $SU(5)$ F-theory GUTs \cite{BHVII, Jiang:2009za, FS5}. Note that this provides a
natural embedding of a flipped GUT in a higher unified structure. It is also possible to consider flux
breaking of $SO(10)$ down to $SU(3) \times U(1) \times SU(2)_{L} \times SU(2)_{R}$ as in \cite{FontIbanI}.
Suitable vevs for additional fields descending from vector-like pairs in the $16 \oplus \overline{16}$ of $SO(10)$
can then break the GUT group down further.

Fluxes also affect the spectrum of zero modes localized on matter curves.
For example, in an $SU(5)$ GUT with matter in the $\overline{5}$, the
decomposition of the $\overline{5}$ reveals that the triplets and doublets couple differently to
the internal hyperflux
\begin{equation}
\overline{5} \rightarrow (\overline{3},1)_{+1/3} \oplus (1,2)_{-1/2}
\end{equation}
In other words, when the net hyperflux through a matter curve is non-zero, we do not
retain full GUT multiplets in the zero mode spectrum. An economical way to evade this
problem is to demand that flux through chiral matter curves is zero \cite{BHVII}. In less
minimal $SO(10)$ models, flux from $U(1)_Y$ and $U(1)_{B-L}$ can also be chosen so that different
components of a GUT multiplet localize on distinct curves \cite{FontIbanI}.

Doublet-triplet splitting in the Higgs sector by fluxes occurs when the Higgs field localizes
on a curve with non-zero hyperflux \cite{BHVII}. Since $H_u$ and $H_d$
have opposite $U(1)_Y$ quantum numbers, this also means that in
order to get a $H_u$ and $H_d$ zero mode, these fields should localize on different matter
curves \cite{BHVII}.

\subsection{Proton Decay}

Proton decay is a signature of GUT models (see \cite{RabyProton} for a review) and has been
studied in F-theory GUT models in \cite{TatarProton, BHVII, DWII, FFAST}. Dangerous
dimension four and five operators which can induce rapid proton decay must be sufficiently suppressed
in order to evade current experimental bounds. On the other hand, the coefficients of dimension six operators
provide a potentially exciting window into GUT physics.

The most phenomenologically problematic couplings are cubic F-terms such as $\overline{5}_{M} \times \overline{5}_{M} \times 10_M$.
Such couplings are absent in the MSSM by assuming the existence of a $\mathbb{Z}_{2}^{matter}$ matter parity
under which chiral matter is odd and the Higgs fields are even. Geometrically, we can exclude
such contributions by demanding the absence of such Yukawa enhancement points.
The absence of such interactions is also compatible with
symmetry considerations, either through embedding $\mathbb{Z}^{matter}_{2} \subset U(1)_{PQ} \subset E_8$ as in \cite{HVGMSB},
or by viewing it as a geometric action of the compactification \cite{BHVII, Tatar:2009jk}. $E_7$ singularity structures in
F-theory GUTs also effectively forbid dimension four proton decay operators \cite{TatarProton}.

Dimension five operators such as
\begin{equation}
\frac{\eta_5}{M_{GUT}} \cdot \int d^2 \theta \, QQQL
\end{equation}
with $\eta_5$ a numerical coefficient, induce rapid proton decay primarily through the
channel $p \rightarrow K^+ \overline{\nu}$. Current bounds on the lifetime
of the proton require $\eta_5 < 10^{-10}$.
In a four-dimensional GUT, doublet-triplet splitting can also inadvertently generate this operator. For example,
when the Higgs up and Higgs down triplets couple to the MSSM
matter fields through interactions such as
\begin{equation}
W_{GUT} \supset T_{u} Q Q + T_d Q L + M_{GUT}T_u T_d,
\end{equation}
heavy triplet exchange will generate the operator $QQQL/M_{GUT}$. This problem is potentially worse
in higher-dimensional models since there is now a whole Kaluza-Klein tower of excitations to worry about.

Doublet-triplet splitting by hyperflux automatically addresses this issue
by geometrically sequestering the Higgs up and down on distinct matter curves. This means that the Higgs
triplets pair up with other Kaluza-Klein modes, and the offending dimension five operator is not
generated \cite{BHVII}. In an intersecting seven-brane configuration, this can also be traced to the presence of
additional anomalous symmetries under which the MSSM fields are charged, such as a $U(1)$ Peccei-Quinn symmetry.
As advocated in \cite{HVGMSB, BHSV, Epoint}, this $U(1)$ naturally fits inside the unfolding of an $E_8$ singularity.

Dimension six operators such as
\begin{equation}\label{decayops}
\frac{\eta_6}{M_{GUT}} \cdot \int d^4 \theta \, U^{\dag} E^{\dag } Q Q
\end{equation}
with $\eta_6$ a numerical coefficient, induce proton decay primarily through the channel $p \rightarrow e^+ \pi^0 $, but
at a level which impinges on the regions currently being probed by experiment.
In a four-dimensional GUT, the coefficient $\eta_{6} \sim g^{2}_{GUT}$, and is generated by
the exchange of off-diagonal gauge bosons of the broken GUT group. A similar effect is present in F-theory GUTs from all of the
off-diagonal elements of the broken gauge group. Moreover, because these modes actually fit inside an entire Kaluza-Klein tower of heavy states,
the full contribution from such effects is summed up by a two-point function integrated over
the internal directions wrapped by the seven-brane. Due to the presence of internal volume factors for the heavy modes participating in the corresponding
Green's function, there is a relative parametric enhancement in some decay channels, by a
factor of $\log \alpha_{GUT}^{-1}$ \cite{WijnholtReview, DWII}. Unfortunately, the overall ambiguity in the normalization of the
internal Green's function means that extracting a precise coefficient is difficult with present techniques
\cite{DWII}.

\subsection{Gauge Coupling Unification}

Starting from the weak scale, and evolving to higher energies, we define the
GUT scale as the energy at which the $SU(2)_L$ and $U(1)_Y$ gauge couplings (normalized
to embed in $SU(5)$) are equal. Unification occurs provided the $SU(3)_C$ gauge
coupling also meets at this same scale. At one loop order,
the MSSM couplings unify, and at two loop order, the value of $\alpha_{3}$ is roughly
$4 \%$ lower than $\alpha_{GUT}$, though the precise value depends on TeV and
GUT scale threshold corrections. Indeed, particular patterns of soft breaking masses can
sometimes help with gauge coupling unification \cite{RabyUNIFY}.

As in many GUT models, the GUT breaking sector itself can shift
the tree level values of the gauge coupling constants. This shift is
comparable in magnitude to the effects from subleading two loop
corrections to the evolution of the MSSM couplings. These contributions
come with various signs, and rather importantly, there exist regimes of parameter space
in F-theory GUTs where gauge coupling unification can be retained \cite{DWII, BlumenhagenUNIFY}. Effects
from the closed string sector connected with hyperflux breaking also affect the running of couplings \cite{Conlon:2009qa},
though a complete discussion of such effects is beyond the scope of the present article.

First consider the contribution from GUT breaking fluxes. In the seven-brane theory, the
gauge field strength couples to itself through kinetic terms, as well as terms quartic
in the field strength. The net contribution is given by the kinetic term for gauge fields of
the eight-dimensional gauge theory, and the Chern-Simons like coupling of the seven-brane
to the background RR zero-form potential $C_0$
\begin{equation}
M_{\ast}^{4}\int_{\mathbb{R}^{3,1} \times S} Tr(F \wedge \ast F) + \int_{\mathbb{R}^{3,1} \times S} C_0 \wedge Tr(F^4)
\end{equation}
where in the above we have suppressed the order one coefficients multiplying each term.
Prior to GUT breaking, the GUT coupling is controlled by the volume wrapped by the seven-brane
so that $\alpha_{GUT}^{-1} = M_{\ast}^{4} \text{Vol}(S)$.
Activating an internal field gauge field strength breaks the GUT group and shifts the relative values
of the gauge coupling constants by
\begin{equation}
\alpha_{i}(M_{GUT})^{-1} \rightarrow \alpha_{i}(M_{GUT})^{-1} + k_{i}
\end{equation}
where $k_i$ is an order one number which depends on the details of the instanton number for
the flux, as well as the breaking pattern. The percentage change in the value of the couplings is
then on the order of $\alpha_{GUT} \sim 5 \%$. In \cite{DWII}, these effects were studied for $U(1)_Y \subset SU(5)$ GUT breaking fluxes, and in \cite{BlumenhagenUNIFY}, a different embedding of abelian fluxes in $U(5)_{GUT}$
was studied in related IIB compactifications. In both cases it was found that the sign of the correction tends to increase the mismatch in unification.

Incomplete GUT multiplets of heavy Kaluza-Klein modes will also induce threshold corrections to unification
\cite{DWII, BlumenhagenUNIFY}. In \cite{BlumenhagenUNIFY}, it was found that
adding a single vector-like pair of Higgs triplets $10^{15} - 10^{16}$ GeV allows unification to be retained.
Summing over the entire tower of Kaluza-Klein states and including the contribution from the $Tr F^{4}$ terms, the specific
linear combination of one loop determinants for the Laplacians of $k-$forms which enters the threshold
correction can be recast as a quasi-topological invariant \cite{DWII, WijnholtComm} known as
the \textit{holomorphic Ray-Singer torsion} \cite{RSTORSION}
\begin{equation}
\mathbb{T} \equiv \frac{1}{2} \sum_{k} (-1)^{k+1} \log \det\,^{\prime }\Delta _{k}
\end{equation}
where $\Delta_{k}$ denotes the Laplacian of bundle-valued $k-$forms,
the one-loop determinants are evaluated in zeta-function regularization,
and $\det^{\prime}$ denotes the determinant with all zero modes omitted. These
threshold effects can in principle have either sign, thus preserving precision
unification in a region of parameter space \cite{DWII}.

\section{Supersymmetry Breaking} \label{SUSYbreak}

So far, our discussion has focussed on the dynamics of the model near the GUT scale. Making contact with observation requires a discussion of supersymmetry breaking, and in particular how all of the superpartners develop masses compatible current experimental bounds. In keeping with the philosophy of local models, in this section we focus on supersymmetry breaking scenarios which rely on dynamics in the vicinity of the seven-brane controlled by the vev of a GUT singlet
\begin{equation}
\langle X \rangle = x + \theta^2 F_X.
\end{equation}
Our aim will be to realize scenarios where the
parameters $F_X$ and $x$ generate viable soft supersymmetry breaking masses for the MSSM gauginos and scalars.

Besides the soft masses, it is also necessary to generate
weak scale coefficients for the $B \mu$ and $\mu$ terms
\begin{equation}
L_{eff} \supset B{\mu} h_u h_d + \int d^{2} \theta \mu H_u H_d + h.c.
\end{equation}
where $h$ denotes the bosonic component of the superfield $H$. The values of these parameters, in addition to the soft supersymmetry breaking
terms in the Higgs sector dictate electro-weak symmetry breaking, and consequently the value of the ratio of the Higgs up and Higgs down vevs, $\tan \beta = v_u / v_d$.

As a brief aside, in many supersymmetric models, especially those derived from string theory,
there are typically many additional nearly flat scalar directions which must also be stabilized.
Here we assume that these moduli have been stabilized through high scale dynamics which will not interfere
with our present discussion. See for example \cite{ConlonBlumenhagen} for recent discussion on
combining certain moduli stabilization scenarios with supersymmetry breaking in F-theory GUT models.

\subsection{Moduli Dominated Scenarios}

Though it is natural to decouple the dynamics of many moduli in a local model, the overall volume modulus of the complex surface
wrapped by the GUT brane remains dynamical, even after decoupling gravity. This suggests using this mode as a source for supersymmetry breaking effects
\cite{LARGE}. Since this modulus couples to all of the MSSM superpartners with particular scaling dimensions, a vev for it
will generate a soft supersymmetry breaking pattern of masses for the MSSM. In \cite{IbanSUSY} the phenomenology
of this scenario was studied in greater detail, where it was found that achieving appropriate electroweak symmetry breaking
requires particular scaling dimensions for the fields of the MSSM. This type of scaling is most easily achieved in configurations where all of
the matter localizes on matter curves. In particular, in such scenarios the Yukawas originate from the triple intersection of seven-branes, which
fits with the main emphasis of models we have presented here.

In this class of models, a bino-like lightest neutralino constitutes
the LSP, providing a natural dark matter candidate for this scenario. Generating
an appropriate relic abundance also requires a stau NLSP, with mass close
to that of the bino, and a large value of $\tan \beta \sim 40$.

\subsection{Gauge Mediation Scenarios}

In a limit where gravity is decoupled from gauge theory, and the volume modulus is stabilized due to high scale dynamics, gauge mediated supersymmetry breaking scenarios are quite natural (see \cite{GiudiceRattazzi} for a review of gauge mediation). Such scenarios are quite attractive, because transmission of supersymmetry breaking by gauge fields does not introduce new sources of flavor violation, which is potentially problematic in other approaches.
Gauge mediated supersymmetry scenarios in F-theory GUTs have been studied in \cite{MarsanoGMSB, HVGMSB}. See for
example \cite{Floratos:2006hs, Cvetic:2008mh} for other work on realizing gauge mediation in string based models.

In minimal gauge mediation, the SUSY breaking GUT singlet $X$ couples to vector-like pairs of messenger fields
$Y \oplus Y^{\prime}$ through the superpotential term
\begin{equation}\label{messint}
W \supset X Y Y^{\prime}.
\end{equation}
Messenger and gauge loops then generate soft supersymmetry breaking masses for the gauginos and scalars of the MSSM.
The precise mass spectrum depends on the scale
of supersymmetry breaking $\sqrt{F_{X}}$, the messenger scale $x$,
the representation content and number of messengers, and the messenger
scale values for the $\mu$ and $B \mu$ terms. As a rule of thumb,
generating TeV scale colored superpartners requires $F_X / x \sim 10^5$ GeV.

This type of scenario has a clean geometric realization in F-theory GUTs \cite{BHVII}.
When $Y$ and $Y^\prime$ localize on distinct matter curves, they will meet a third curve,
normal to the GUT seven-brane, on which $X$ localizes. In principle, $Y$ and $Y^\prime$ can correspond to matter in the
$5 \oplus \overline{5}$ or the $10 \oplus \overline{10}$.

Gauge mediated interactions generate soft masses, but do not address how the $\mu$ term of the MSSM is generated. The first issue is to avoid a GUT scale value for the $\mu$ parameter in the MSSM superpotential
\begin{equation}
W_{MSSM} \supset \mu H_u H_d.
\end{equation}
This condition is met in models where the Higgs up and down are sequestered on distinct matter curves, because the requisite term
will not be invariant under the gauge symmetries of the other seven-branes intersecting the GUT stack \cite{BHVII, HVGMSB}.
In the low-energy effective theory, this corresponds to a nearly exact anomalous $U(1)$ Peccei-Quinn symmetry. Such a PQ symmetry
descends from another seven-brane of the compactification which intersects the GUT stack so that all matter fields are
charged under $U(1)_{PQ}$. As a point of terminology, we shall refer to this as the \textit{PQ seven-brane} of the compactification.

In minimal setups achieving a weak scale $\mu$-term determines the required scale of supersymmetry breaking. The higher dimension operator
\begin{equation}
\int d^4 \theta \frac{X^{\dag} H_u H_d}{\Lambda_{UV}}
\end{equation}
induces $\mu \sim F_X/\Lambda_{UV}$. In the geometry, this operator is generated at points where the $X$ curve and Higgs curves meet
\cite{HVGMSB}. The scale $\Lambda_{UV} \sim 10^{15}$ GeV is close to the GUT scale \cite{BHVII, HVGMSB}, which requires
$F_X \sim 10^{17}$ GeV$^2$ to generate a weak scale $\mu \sim 100$ GeV. Combined with the requirement $F_X / x \sim 10^5$ GeV,
this implies $x \sim 10^{12}$ GeV \cite{HVGMSB}.

The analogous operator compatible with $U(1)_{PQ}$ which generates the $B \mu$ term is
\begin{equation}
\int d^4 \theta \frac{X^{\dag} X X^{\dag} H_u H_d}{\Lambda_{UV}^{3}}
\end{equation}
producing a value for $B \mu \sim (x/\Lambda_{UV}) \cdot \mu^2$, which is far smaller than $\mu^2$. This is the primary contribution to the
$B \mu$ term because of localization of the matter fields on Riemann surfaces. This yields the messenger scale
boundary condition $B \mu \sim 0$. In addition, the messenger scale $A$-terms are also zero due to brane localization, so that
additional potentially problematic CP violating phases are not generated. Finally, at low energies, this
scenario leads to large values of $\tan \beta \sim 20 - 35$.

Additional contributions to the soft masses will arise from anomalous $U(1)$ gauge bosons which couple the supersymmetry breaking
sector to the visible sector \cite{ArkaniHamed:1998nu}. In \cite{HVGMSB}, the effects of integrating out an anomalous $U(1)_{PQ}$ gauge
boson were also studied, where it was shown that there are additional contributions
to the soft breaking parameters from higher dimension operators of the form
\begin{equation}
L_{eff} \supset - g^{2}_{PQ} e_X e_\Psi \int d^4 \theta \frac{X^{\dag} X \Psi^{\dag} \Psi}{M_{U(1)_{PQ}}^2}
\end{equation}
where $e_X$ and $e_\Psi$ denote the PQ charges of $X$ and $\Psi$, $g_{PQ}$ denotes the gauge coupling of the $U(1)_{PQ}$ gauge theory,
and $M_{U(1)_{PQ}}$ is the mass of the $U(1)_{PQ}$ gauge boson. In principle $M_{U(1)_{PQ}}$
can range from masses somewhat below the GUT scale to much higher values.

Combining the effects of minimal gauge mediation with the effects of heavy PQ
gauge boson exchange induces a \textit{PQ deformation} of the soft masses away from the usual minimal gauge mediation spectrum
\begin{equation}
m_{soft}^{2} = m^{2}_{mGMSB} - q \Delta_{PQ}^{2}
\end{equation}
where $\Delta_{PQ} \sim g^2_{PQ} F/M_{U(1)_{PQ}}$, and $q \sim - e_X e_\Psi$ is determined by the product of the $X$ and $\Psi$ PQ charge of the field $\Psi$ in question. The PQ charges of the MSSM chiral matter is opposite in sign to that of $X$. This has the effect of lowering the mass of the scalars, such as the stau, but does not shift the masses of the gauginos. The actual value $q$ depends on the neutrino sector of the model. This is because compatibility with the other interaction terms of the MSSM uniquely determines a choice of PQ charge assignments compatible with the neutrino scenarios of \cite{BHSV}. See figure \ref{egut} for a plot of the characteristic mass spectrum of a Majorana neutrino scenario at minimal and maximal PQ deformation.

\begin{figure}[ptb]
\centerline{\psfig{figure=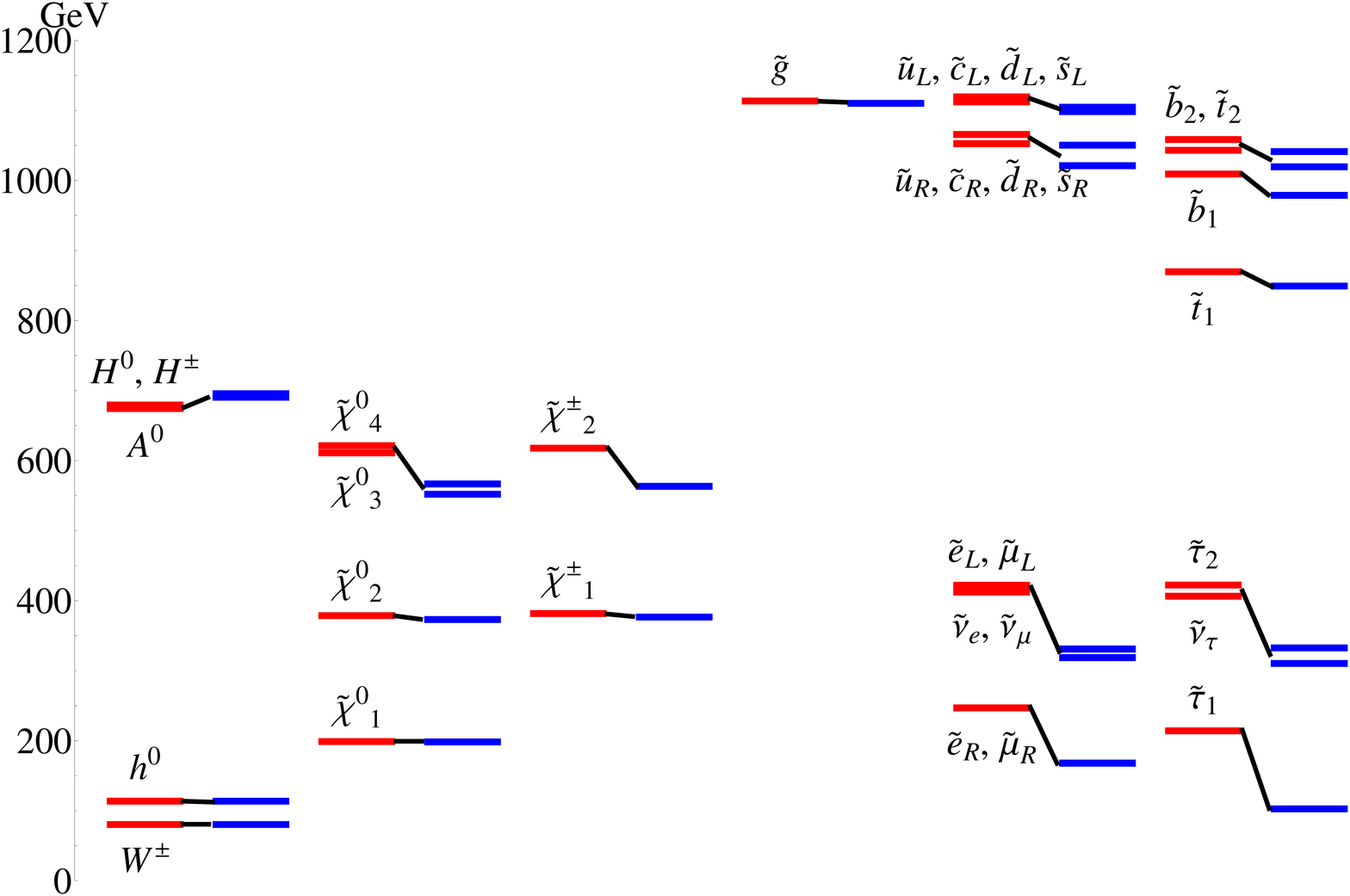,height=10cm}}
\caption{Plot of the sparticle mass spectrum (indicated by
$\widetilde{}$'s) in an F-theory GUT with one vector-like
pair of messengers in the $10 \oplus \overline{10}$ and
$F_X / x = 5 \times 10^4$ GeV for minimal (left, red columns) and maximal
(right, blue columns) PQ deformation. At $\Delta^{min}_{PQ} = 0$, a bino-like
lightest neutralino $\widetilde{\chi}^{0}_1$ is the NLSP. For moderate values of $\Delta_{PQ} \sim 100$ GeV, the NLSP transitions
to the lightest stau $\widetilde{\tau}_1$, which persists until a tachyon develops near $\Delta^{max}_{PQ} \sim 200$ GeV.}
\label{egut}%
\end{figure}

Generating a hierarchically suppressed supersymmetry breaking scale is more challenging,
but can be arranged through a Polonyi term
\begin{equation}
W_{eff} \supset \exp(-\text{Vol}(S_{PQ})) X
\end{equation}
induced by Euclidean D3-instantons wrapped on the PQ seven-brane
\cite{HMSSV, MarsanoTool, MarsanoGMSB, HVGMSB}. Here, realizing the precise
scales $x \sim 10^{12}$ GeV, $F_{X} \sim 10^{17}$ GeV$^{2}$ requires a certain
amount of tuning in the K\"ahler potentials of $X$ and other axion-like fields which
contribute to the QCD axion \cite{HVGMSB}. In principle, it is also possible to use
the supergravity effective potential to generate an appropriate potential for
$X$ \cite{SweetSpot, MarsanoGMSB}, though this requires additional assumptions about
the cosmological constant.

The supersymmetry breaking sector also feeds into the physics of axions. The phase of $X$ corresponds to a Goldstone mode
of the nearly exact global $U(1)_{PQ}$ symmetry,
and constitutes along with other axion-like fields the QCD axion \cite{HVGMSB}. Interestingly,
the numerology demanded by gauge mediation that
$x \sim 10^{12}$ GeV is consistent with the available axion window. The axion supermultiplet also influences the cosmology of F-theory GUTs \cite{FGUTSCosmo}. The bosonic partner of the axion -- the saxion -- develops a weak scale mass due to supersymmetry breaking. The fermionic component corresponds to the axino,
which is also the spin 1/2 component of the gravitino. As the Goldstino of local supersymmetry breaking, the gravitino has a
mass of $F_{X}/M_{pl} \sim 10 - 100$ MeV set by the supersymmetry breaking scale. The saxion
field oscillates in the early Universe, eventually decaying prior to the start of big bang nucleosynthesis. When it decays, it releases a significant amount of entropy. The diluted relic abundance of gravitinos makes it a viable dark matter candidate \cite{FGUTSCosmo}. In principle, axionic dark matter can also comprise a component of dark matter. Note that saxion decay also dilutes the relic abundance of other possible
dark matter candidates \cite{Epoint}.

\section{Flavor in F-theory} \label{FlavorFth}

Proceeding down in energy scales, we have studied the physics at the GUT scale, and supersymmetry breaking. We now turn to realizations of
flavor for the quarks and leptons of the Standard Model, and its extension to massive neutrinos.

The masses and mixing angles of the quark sector exhibit striking and rather mysterious flavor hierarchies.
For example, the mass of the $u,c,t$ quarks are respectively $0.002, 1, 170$ GeV, and the magnitudes of the CKM quark
mixing matrix elements are \cite{PDG}
\begin{equation}\label{CKMexp}
|V_{CKM}| \sim
\left(
\begin{array}
[c]{ccc}%
|V_{ud}| & |V_{us}| & |V_{ub}| \\
|V_{cd}| & |V_{cs}| & |V_{cb}| \\
|V_{td}| & |V_{ts}| & |V_{tb}|%
\end{array}
\right)
\sim
\left(
\begin{array}
[c]{ccc}%
1 & 0.2 & 0.004\\
0.2 & 1 & 0.04\\
0.008 & 0.04 & 1%
\end{array}
\right).
\end{equation}
By contrast, neutrino oscillation experiments indicate a far lower characteristic mass scale of
order $0.05$ eV, and compared to the CKM matrix, the leptonic mixing matrix $V_{PMNS}$ is less hierarchical \cite{PDG}.
Moreover, the magnitude of the $(1,3)$ element of the PMNS matrix
$|V^{1,3}_{PMNS}| = \sin \theta_{13}$ is still consistent with a value of zero,
and is bounded above by $0.2$.

GUT structures naturally explain some aspects of flavor.
For example, the mass of the $b$ quark and $\tau$ lepton unify at the GUT scale \cite{BTAUunify},
which fits with embedding their Yukawas in the interaction term
$\overline{5}_H \times \overline{5}_M \times 10_M$. Geometrically, order one coefficients for both the
$\overline{5}_H \times \overline{5}_M \times 10_M$ and
$5_H \times 10_M \times 10_M$ seem the easiest to arrange,
and this is the case we focus on here. Since the top and bottom quark have different masses,
this is most compatible with large $\tan \beta$ scenarios.

The limited representation content for zero modes in an F-theory GUT
eliminates some common mechanisms in four-dimensional models \cite{GeorgiJarlskog} for correlating GUT
breaking with flavor structure. On the other hand, these
same geometric ingredients suggest new possibilities, both for quarks and leptons.
To frame our discussion, we again focus on minimal $SU(5)$ F-theory GUTs.
We show that minimal geometries required to realize the MSSM superpotential
also produce rank one Yukawas. Subleading corrections to these
Yukawas responsible for flavor hierarchies are then induced through violations of local symmetries,
either from background ambient three-form fluxes in the case of quarks,
or from non-chiral massive modes of the compactification, in the case of neutrinos.

\subsection{Geometry of Flavor}

The overlap of matter field wave functions in the internal
directions of the seven-brane determine the up-type Yukawa matrix
\begin{equation}
\lambda^u_{ij} = \int_S \psi^{H_u} \psi^Q_i \psi^U_j
\end{equation}
with similar expressions for the down-type and charged lepton Yukawas. The maximal overlap
between matter localized on different curves occurs where these curves geometrically intersect. To leading order,
the up-type Yukawa matrix is given by a sum over all such intersection points
\begin{equation}
\lambda^{u}_{ij} = \sum_{p} \psi^{H_u}(p) \psi^Q_i(p) \psi^U_j(p).
\end{equation}
Note that each summand is an outer product
of two flavor vectors, and so defines a rank one $3 \times 3$ matrix. In other words,
a single interaction point already generates the expected structure of one generation
of weak scale mass, and two generations which to first approximation are massless
\cite{BHVII, HVCKM}. In actual compactifications there will be additional enhancement points of the
geometry \cite{WatariPoints, Cordova}. Realizing a single heavy generation then suggests that either
these points are tuned to be close together \cite{BHVII, WatariPoints}, or that the MSSM fields
only participate in one such point \cite{BHVII, HVCKM}.

The number of heavy generations depends on how many curves participate in a Yukawa interaction.
To see this, let us consider a model where the three generations of matter are distributed on
up to three matter curves. The $5 \times 10 \times 10$
interaction requires a point of enhancement from $SU(5)$ to at least $E_6$. Returning to our discussion near
equation \eqref{tripinteraction}, this interaction term couples two $10$'s which would seem
to localize on different matter curves since they have distinct charges under the abelian factor of
$SU(5) \times U(1)^{2} \subset E_6$. This is phenomenologically problematic
because the Yukawa matrix will then have zeroes along the diagonal. Assuming
roughly similar normalizations for the kinetic terms, this corresponds to a
model with only one hierarchically light generation \cite{BHVII}.

In actual compactifications of F-theory the
curves on which the two $10$'s localize are often just different branches
of a single complex one-dimensional curve \cite{tataretal}. This is an example of a more general phenomenon
known as \textit{seven-brane monodromy}, and constitutes an important element in
achieving realistic flavor structure in an F-theory GUT.

\subsubsection{Seven-Brane Monodromy}

Seven-brane monodromy is best motivated by example. To this end consider
the breaking pattern defined by unfolding an $A_{n-1}$ singularity $y^2 = x^2 + z^{n}$ as
\begin{equation}
y^2 = x^2 + z^{n-2}(z^2 + 2\beta z + \gamma) = x^2 + z^{n-2}(z - t_+)(z - t_-),
\end{equation}
where $t_{\pm} = - \beta \pm \sqrt{\beta^2 - \gamma}$.
This corresponds to an $SU(n-2)$ stack of seven-branes at $z = 0$,
and another seven-brane at $z^2 + 2\beta z + \gamma = 0$. The branch cuts
in $t_{\pm}$ identifies the two seemingly different seven-branes at $z = t_+$ and $z = t_-$.

Returning to our discussion near equation \eqref{tripinteraction}, in terms of the eight-dimensional $SU(n)$ gauge theory locally defined at $z = t_{\pm} = 0$, the vevs of the Casimirs of $\Phi$ can be packaged in terms of the coefficients in the unfolding of the singularity \cite{BHVI}. When
$\Phi$ is not diagonalizable, the corresponding matter curves exhibit branch cuts, reflected in the $t_{\pm}$.

We now define seven-brane monodromy in more general terms. Consider again a configuration where the singularity type is $G_S$ at
generic points of the complex surface wrapped by the seven-brane, which enhances to $G_p$ at a point of the geometry.
The unfolding of this singularity is specified by coordinates $t_i$ of the Cartan subalgebra of $G_p$,
which are to be thought of as the eigenvalues of $\Phi$. The
\textit{monodromy group} of the seven-brane configuration $G_{mono}$ permutes the $t_{i}$'s. In a
breaking pattern of the form $G_{p} \supset G_{S} \times \Gamma$, $G_{mono}$ is a
subgroup of the Weyl group of $\Gamma$.

Returning to the example of the $5 \times 10 \times 10$ interaction, note that under
the breaking pattern $E_{6} \supset SU(6) \times SU(2) \supset SU(5) \times U(1) \times SU(2)$, the adjoint of $E_6$ decomposes as
\begin{equation}
78 \rightarrow (24,1) \oplus (1,3) \oplus (5_{-6},1) \oplus (\overline{5}_{+6},1) \oplus (10_{-3},2) \oplus (\overline{10}_{-3},2)
\end{equation}
so that the $10$'s transform as a doublet of $SU(2)$. Indeed, the Weyl group of $SU(2)$ is $\mathbb{Z}_{2}$,
and monodromy acts by interchanging the components of the doublet. The geometry then consists of a single $E_6$
enhancement point where one $10$ and one $5$ curve meet to form the $5 \times 10 \times 10$ interaction. Thus, monodromy in the seven-brane configuration
allows us to achieve two hierarchically light generations \cite{tataretal}.

\subsection{Quark Yukawas}

We now turn to the Yukawas of the lighter quark generations. One common way to
generate flavor hierarchies is through the Froggatt-Nielsen mechanism \cite{Froggatt:1978nt}.
In this scenario, the matter fields of the Standard Model are charged under an
additional horizontal symmetry, where different generations have distinct Froggatt-Nielsen charges.
Spontaneous breaking of this horizontal symmetry generates additional corrections
$\lambda_{ij} \sim \varepsilon^{a_i + b_j}$, producing hierarchical masses and mixing angles. This begs the question, however, as
to how the parameters $a_i$ and $b_j$ are to be chosen, as well as how the parameter $\varepsilon$ is fixed.

It is possible to engineer F-theory GUTs utilizing this four-dimensional mechanism \cite{Dudas:2009hu}. In fact, even
without introducing explicit GUT singlets, the ingredients of an F-theory compactification provide similar, though more stringy
variants on this theme. Matter localized on complex one-dimensional
curves have similar charges induced by the action of the internal Lorentz group \cite{HVCKM}. More precisely, the matter wave functions
satisfy $\overline{\partial}_A \psi = 0$, and so are locally holomorphic. Organizing the wave function
solutions according to their order of vanishing near a Yukawa point, we have $\psi^i \sim z^{3-i} + O(z^{4-i})$
for $i=1,2,3,$ in a three generation model. The action of the internal Lorentz group corresponds to a rephasing symmetry of $z$,
providing a horizontal symmetry of the chiral generations. Background gauge field and three-form fluxes of an F-theory
compactification distort the profile of matter field wave functions \cite{HVCKM}
\begin{equation}\label{expequation}
\psi \rightarrow \exp(\mathcal{M}_{i \overline{j}} z_i \overline{z}_{\overline{j}}) \psi
\end{equation}
where the $z_i$ denote local holomorphic coordinates in a patch of the Yukawa point. This distortion of wave functions then suggests a possible
mechanism for generating corrections to Yukawas. Note that by dimensional analysis, $\mathcal{M}$ scales as $M_{GUT}^{2}$.

Even so, the topological structure of a single non-zero entry turns out to be quite robust against
physical perturbations. Gauge field fluxes, for example, distort the profile of the
physical matter field wave functions, but fail to distort the Yukawa
structure \cite{FGUTSNC, ConlonPalti}. Yukawa distorting fluxes correspond
to three-form fluxes such as $H = H_R + \tau H_{NS}$, which induce
higher dimension operator deformations of the seven-brane gauge theory. The effects of
three-form $H$-flux are locally captured in a patch of the Yukawa enhancement point in
terms of a non-commutative deformation of the seven-brane theory \cite{FGUTSNC}.
In principle, instanton effects can also generate similar non-commutative
deformations \cite{Marchesano:2009rz}, the effects of which can also be captured in terms
of an effective contribution from $H$-fluxes \cite{Baumann:2009qx}.

Having different hypercharges, the different components of a GUT multiplet will couple to the background fluxes
differently, leading to different wave function and Yukawa distortions for the various generations. This is quite
important, because while the bottom quark and $\tau$ lepton mass unify
near the GUT scale, the masses of other particles $m_c \neq m_{\mu}$ and $m_{d} \neq m_{e}$ do not.
In other words, the geometry of the compactification preserves a leading order notion of matter unification, which is violated by
flux distortion of the wave functions \cite{BHVII, HVCKM, FGUTSNC}.

Expanding in successive powers of $\overline{z}$ in $\mathcal{M}$, either by expanding in higher powers of
$\mathcal{M}$ in the exponential of equation \eqref{expequation}, or by including
higher order derivative corrections yields the structures
\cite{HVCKM}
\begin{equation}
\lambda_{DER} \sim \left(
\begin{array}
[c]{ccc}%
\varepsilon^{5} & \varepsilon^{4} & \varepsilon^{3}\\
\varepsilon^{4} & \varepsilon^{3} & \varepsilon^{2}\\
\varepsilon^{3} & \varepsilon^{2} & 1%
\end{array}
\right) \text{,}\,\,\, \lambda_{FLX} \sim \left(
\begin{array}
[c]{ccc}%
\varepsilon^{8} & \varepsilon^{6} & \varepsilon^{4}\\
\varepsilon^{6} & \varepsilon^{4} & \varepsilon^{2}\\
\varepsilon^{4} & \varepsilon^{2} & 1
\end{array}
\right)
\end{equation}
where each entry is multiplied by an order one complex coefficient which we suppress. As the name suggests, the $\lambda_{DER}$ expansion is given by expanding to first order in the flux and then expanding in successive gradients or derivatives of the flux. The $\lambda_{FLX}$ expansion is given by expanding to leading order in the gradients of the flux, and then expanding in successive powers of the first gradient.

Because the matter field wave functions couple differently to the background fluxes, the precise value of $\varepsilon$ and
the dominance of the FLX or DER expansion can depend on the matter field in question \cite{HVCKM}. Identifying up-type quark
Yukawas with $\lambda_{FLX}$ and down-type quarks
with $\lambda_{DER}$, this yields hierarchical up-type quark masses
\begin{equation}
m_{u} : m_{c} : m_{t} \sim \varepsilon_u^{8} : \varepsilon_u^{4} : 1
\end{equation}
and down-type quark masses
\begin{equation}
m_{d} : m_s : m_b \sim \varepsilon_d^{5} : \varepsilon_d^{3} : 1.
\end{equation}
Though this is somewhat heuristic, it reproduces the observed masses
and mixing angles surprisingly well \cite{HVCKM}. Numerically,
achieving a match to the observed quark masses works best when
$\varepsilon_{u} \sim 0.26$ and $\varepsilon_d \sim 0.27$. This is quite close to the
value expected based on the relative scaling of $\mathcal{M}$ to the string scale
$\varepsilon \sim M_{GUT}^{2} / M_{\ast}^2 \sim \sqrt{\alpha_{GUT}} \sim 0.2$ \cite{HVCKM, FGUTSNC}.

Achieving a hierarchical CKM matrix is more subtle. The CKM matrix measures the mismatch between
the gauge and mass quark eigenstates. Since the profile of matter field wave functions varies over
the geometry of the seven-brane, achieving a roughly diagonal CKM matrix requires near alignment of
the up and down type Yukawa points. Assuming
that this further condition for point unification has been met,
the resulting CKM matrix is then \cite{HVCKM, FGUTSNC}
\begin{equation}
V^{F-th}_{CKM} \sim \left(
\begin{array}
[c]{ccc}%
1 & \varepsilon & \varepsilon^{3}\\
\varepsilon & 1 & \varepsilon^{2}\\
\varepsilon^{3} & \varepsilon^{2} & 1%
\end{array}
\right) \sim \left(
\begin{array}
[c]{ccc}%
1 & 0.2 & 0.008\\
0.2 & 1 & 0.04\\
0.008 & 0.04 & 1%
\end{array}
\right)
\end{equation}
at the GUT scale. In principle, this should be evolved to lower energy
scales, but this turns out to multiply various entries by order one
coefficients, which is already beyond the approximation scheme adopted here.
Returning to equation \eqref{CKMexp}, it is surprising how well these crude
numerical estimates match with observation.

From the perspective of Yukawa enhancements, putting $E_6$ and $SO(12)$ together
can be accomplished through a higher unification to either
$E_7$ or $E_8$. Including neutrinos pushes this further
to $E_8$ \cite{Epoint}. We will return to this theme later when we discuss
minimal unification at a point of $E_8$ in section \ref{eunif}.

\subsection{Neutrino Models}

The numerology of the seesaw mechanism $m_{\nu} \sim M_{weak}^2/\Lambda_{UV}$, for $\Lambda_{UV}$ close to
the GUT scale strongly suggests a connection between GUTs and neutrinos. It is well known that
Majorana and Dirac neutrino masses can respectively be generated by the higher dimension operators
\begin{equation}\label{HigherOps}
\int d^2 \theta \frac{(H_u L)^2}{\Lambda_{UV}}, \text{\,\,\,}
\int d^4 \theta \frac{H_{d}^{\dag} L N_{R}}{\Lambda_{UV}}.
\end{equation}
The Higgs vevs $H_u \sim M_{weak}$
and $H_d \sim \theta^2 F_{H_{d}} \sim \theta^2 M_{weak}^2$ induce the mass terms
\begin{equation}
\int d^2 \theta \frac{M_{weak}^2}{\Lambda_{UV}} N_L N_L, \text{\,\,\,}
\int d^2 \theta \frac{M_{weak}^2}{\Lambda_{UV}} N_L N_R.
\end{equation}

Right-handed singlets which interact with the MSSM fields can originate from matter
localized on curves normal to the GUT stack \cite{BHVII, RandDuff, BHSV}, but
can also in principle originate from moduli fields \cite{Tatar:2009jk}. Right-handed neutrinos in the spinor $16$ of an
$SO(10) \supset$ flipped $SU(5)$ F-theory GUT are also possible \cite{BHVII, Jiang:2009za, FS5}.

The flavor structure of $SU(5)$ models with neutrinos localized on singlet curves has been developed more,
and so we focus on this case. In such scenarios, the interaction with the MSSM fields localizes at a point
of enhancement from $SU(5)$ to at least $SU(7)$. Decomposing the adjoint of $SU(7)$ to $SU(5) \times U(1)^2$ yields
\begin{equation}
48 \rightarrow 24_{0,0} \oplus 1_{0,0} \oplus 1_{-2,0} \oplus 1_{2,0} \oplus 5_{+1,-7}
\oplus 5_{-1,-7} \oplus \overline{5}_{-1,+7} \oplus \overline{5}_{+1,+7}
\end{equation}
so that the MSSM fields correspond to a pairing of the $5$ and $\overline{5}$ with a GUT singlet $1_{\pm 2, 0}$,
corresponding to the right-handed neutrino. Depending on the $U(1)$ charge assignments of the matter fields
and the neutrino scenario in question, integrating out massive modes of
the compactification will then generate the higher dimension
operators of equation \eqref{HigherOps}. For Majorana scenarios, seven-brane monodromy
is especially important because it breaks $U(1)$ symmetries which would
otherwise forbid Majorana mass terms.

The participation of massive non-chiral modes dilutes
the flavor hierarchies in comparison to what is present
in the quark and charged lepton Yukawas yielding neutrino masses
\begin{equation}
m_{\nu_1} : m_{\nu_2} : m_{\nu_3} \sim \varepsilon^{2} : \varepsilon : 1,
\end{equation}
which is known as a normal mass hierarchy scenario. Up to order one factors,
the predicted ratio of atmospheric and solar neutrino mass splittings is then
\cite{HKSV}
\begin{equation}
\frac{\Delta m^2_{sol}}{\Delta m^2_{atm}} \sim \varepsilon^2 \sim
\alpha_{GUT} \sim 0.04
\end{equation}
which is numerically quite close to the experimental value of
$0.03$ \cite{PDG}. Again, let us stress that this numerology works surprisingly
well considering all of the crude approximations performed.

Large neutrino mixing angles will generically be present in models where the
neutrino interaction point is geometrically separated from the lepton interaction
point. This is in accord with the observed large mixing angles $\theta_{atm}$ and
$\theta_{sol}$, but leads to a certain amount of tension with the current upper
bound on the mixing angle $\theta_{13} < 0.2$.

As in the quark sector, realizing a hierarchical mixing matrix naturally suggests unifying the charged lepton and
neutrino interaction points. Even when the neutrino and
charged lepton interaction points are close together, the
Kaluza-Klein dilution of flavor hierarchies still generates large mixing angles with
PMNS mixing matrix \cite{BHSV}
\begin{equation}
V_{PMNS} \sim \left( \begin{array}
[c]{ccc}%
V_{e 1} & \varepsilon^{1/2} & \varepsilon\\
\varepsilon^{1/2} & V_{\mu 2} & \varepsilon^{1/2}\\
\varepsilon & \varepsilon^{1/2} & V_{\tau 3}%
\end{array}
\right)
\end{equation}
where the $V$'s are constrained by unitarity of $V_{PMNS}$. Using the same parametric scaling $\varepsilon \sim \sqrt{\alpha_{GUT}}$
as in the quark sector yields $\varepsilon^{1/2} \sim 0.45$, leading to order one
solar and atmospheric neutrino mixing. In addition, the model also predicts
\begin{equation}
\theta_{13} \sim \alpha_{GUT}^{1/2} \sim 0.2.
\end{equation}
In other words, the expectation is
that $\theta_{13}$ is close to the current experimental bound. This appears
to also be a common feature of neutrino models with additional interaction points
\cite{RandDuff}.

Majorana and Dirac neutrino scenarios can in principle be distinguished
by neutrinoless double beta decay experiments. The corresponding decay amplitude
is only generated in Majorana scenarios, and is proportional to a mass term
$m_{\beta \beta}$, which in F-theory GUTs is on the order of $6$ meV \cite{BHSV, RandDuff}.
This value is quite small, and may only be within reach of experiments on the
horizon of a decade \cite{EXO}.

More options are available in less minimal scenarios. Anarchic neutrino
mixing and mass matrices can be realized either by separating the
charged lepton and neutrino interaction points \cite{BHSV}, or by including multiple
neutrino enhancement points \cite{RandDuff}. Tuning the locations of the
neutrino interaction points, neutrino models with an inverted mass hierarchy
can also be realized \cite{RandDuff}.

\section{Minimal $E_8$ Unification}\label{eunif}

In this article, we have demanded that the geometry of an F-theory GUT satisfy a long list of phenomenological constraints. These include
a list of matter curves where GUT multiplets localize, interaction terms for the MSSM superpotential, neutrino sector,
and supersymmetry breaking sector (for gauge mediation models). From this perspective, each requirement would seem to add an additional
degree of arbitrariness to the models we have considered. In addition, it is not altogether clear whether all of these ingredients can
in fact consistently combine in a single, unified framework.

The requirements of flavor physics that the quark and leptons exhibit hierarchies in the
mixing matrices suggests unifying the $E_6$, $SO(12)$ and $SU(7)$
interaction points into a single point of $E_8$ enhancement \cite{BHSV, Epoint}.
This $E_8$ enhancement breaks down to $SU(6)$ and $SO(10)$ along the GUT matter curves,
and is also compatible with the supersymmetry breaking sector introduced in section \ref{SUSYbreak}.
The matter curves and interaction terms are then determined by the breaking pattern for
the adjoint of $E_8 \supset SU(5)_{GUT} \times SU(5)_{\bot}$
\begin{equation}
248 \rightarrow (24,1) \oplus (1,24) \oplus (5,10) \oplus (\overline{5},\overline{10}) \oplus (\overline{5},10) \oplus (5,\overline{10}).
\end{equation}
Given the large set of independent directions in the Cartan of $SU(5)_{\bot}$, such an unfolding might appear to generate
a large number of extraneous matter curves. Seven-brane monodromy, which already figures
prominently in the physics of flavor propagates to other sectors of the model, constraining the ways
that additional matter can be added. In fact, it is possible to classify the possible breaking patterns of $E_8$ with seven-brane monodromy,
subject to the physical conditions
\begin{itemize}
\item{Hierarchical CKM and PMNS matrix}
\item{Kaluza-Klein seesaw for neutrinos}
\item{$U(1)_{PQ}$ to forbid a bare $\mu$ term}
\item{$\mu$ term from either $\int d^{4} \theta \frac{X^{\dag} H_u H_d}{\Lambda_{UV}}$ or $\int d^2 \theta S H_u H_d$}
\end{itemize}
where $X$ and $S$ are to be thought of as GUT singlets which develop suitable vevs to induce a $\mu$ term. In addition to identifying many matter curves,
the monodromy group also identifies the $U(1)$'s in $SU(5)_{\bot}$, so that only $U(1)_{PQ}$ survives for Majorana neutrino scenarios, and in Dirac neutrino scenarios, only $U(1)_{PQ}$ and $U(1)_{B-L}$ survive. See figure \ref{egut} for a depiction of a minimal $E_8$ model.

\begin{figure}[ptb]
\centerline{\psfig{figure=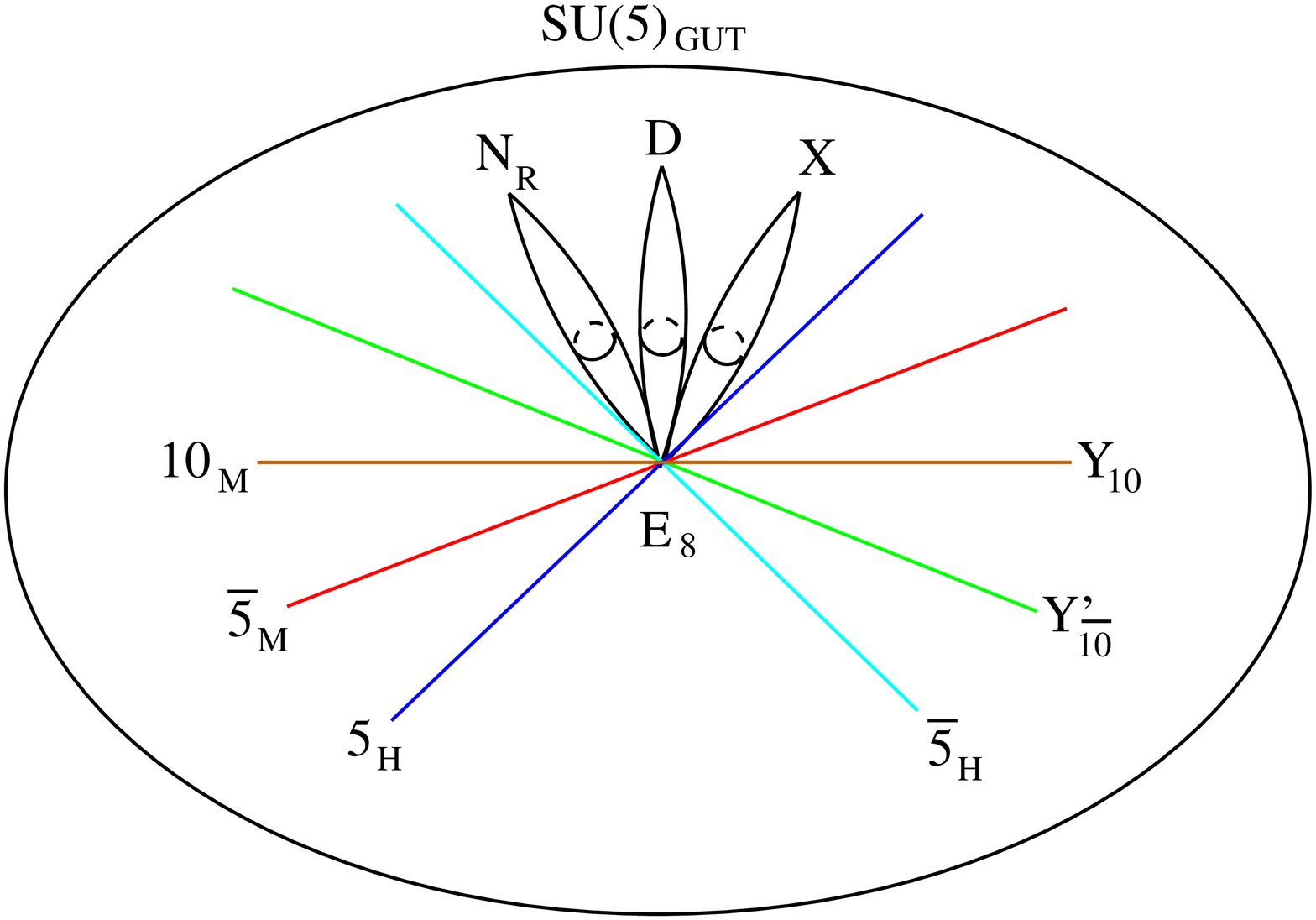,height=8cm}}
\caption{Depiction of a minimal model with $E_{8}$ point unification. In the
proposed model, all of the interaction terms descend from a single $E_{8}$
point of enhancement. Here, matter curves inside the seven-brane are depicted
by straight lines and curves normal to the seven-brane are depicted by
cigar-shaped tubes. In these minimal models, there is typically just enough
room to accommodate the MSSM\ spectrum, and a minimal messenger sector ($Y$'s)
in the $10\oplus\overline{10}$.}
\label{egut}%
\end{figure}

There are typically just enough matter curves to accommodate the Standard Model, and a very constrained messenger sector. In all Majorana neutrino scenarios, and all but one Dirac neutrino scenario, the messengers transform in vector-like pairs in the $10 \oplus \overline{10}$. Moreover, the matter curve for the $10_M$ also supports the messenger \cite{Epoint}. In loop exchange diagrams involving such messenger fields, increasing the number of messengers or equivalently the dimension of the representation tends to lower the masses of the scalar partners relative to the gauginos. In particular, once the effects of the PQ deformation are included, the mass of the lightest stau tends to be lowered relative to the mass of the bino \cite{Epoint}.

\subsection{Experimental Signatures}

The qualitative experimental signatures of this minimal gauge mediation model are dictated by the scale of
supersymmetry breaking, and the identity of the NLSP. The NLSP decays to its Standard Model counterpart and the gravitino with decay rate
$F_X^2/m^{5}_{NLSP}$, leading to a lifetime on the order of one second to an hour. This means that on timescales probed
by colliders, the NLSP is quasi-stable, which is somewhat different from other gauge mediation scenarios.

In bino NLSP scenarios, the characteristic missing energy plus two prompt photon
signature of low scale gauge mediation models is not reproduced here. Rather, the bino will leave the detector as missing energy, much as in gravity mediated supersymmetry breaking. Nevertheless, the different particle spectrum, especially for colored states allows such scenarios to be distinguished from F-theory GUTs \cite{HKSV}.

A characteristic feature of quasi-stable stau NLSP scenarios is that once produced in a collider, it will appear as a heavy charged particle which
registers in a detector as a ``fake muon''. Compared to the signatures of other supersymmetric models with a bino LSP or quasi-stable NLSP, this is a clean signature with low Standard Model background. The expected collider phenomenology of such F-theory GUT scenarios at the CERN Large Hadron Collider has been 
studied in \cite{Heckman:2010xz}.

\section{Conclusions} \label{Conclude}

F-theory GUTs provide a geometric framework for connecting string scale physics to phenomena of the Standard Model. In this article we have
introduced the primary ingredients which enter into such constructions, emphasizing the tight
interplay between these ingredients in phenomenological models. While flexible enough to
accommodate many aspects of the Standard Model, the geometry is rigid enough to favor
particular phenomenological scenarios.

Minimal scenarios which make use of only ingredients found within a
single $E_8$ point seem particularly rigid and predictive. What is particularly interesting is
that within this specific class of models, there are various avenues by which
these models can be falsified and thus indirectly tested by the LHC, and other upcoming experiments for
neutrinos, dark matter and possibly proton decay.

The construction of explicit geometries realizing all of these ingredients remains an active area of
investigation, and such considerations will likely provide an important guide
for future model building efforts. Consistently coupling such models to gravity remains an important
avenue of investigation, potentially providing new constraints which cannot be seen in purely local models.
Moreover, extracting the collider and flavor physics signatures of broader classes
of such vacua may help to illuminate whether F-theory GUTs are realized in Nature.

While any one aspect of a model could be viewed as suggestive, the fact that constraints from the geometry
propagate to many aspects of an F-theory GUT makes the prospects of non-trivial correlations and the prospect of contact with
distinct experiments especially exciting.

\vskip 20pt

{\bf Acknowledgements.}

We thank C. Vafa for a very stimulating collaboration which led
to much of the work reviewed here. In addition,
we also thank our collaborators on related F-theory projects. We also also thank
J. Marsano, S. Raby, C. Vafa, B. Wecht and E. Witten for helpful comments on the draft.
The research of JJH is supported by NSF grant PHY-0503584.

\bibliographystyle{arnuke_revised}

\end{document}